\documentclass[a4paper,11pt,headings=big]{article}
\pdfoutput=1

\usepackage{amsmath,amssymb,graphicx,multirow,xspace}
\usepackage[colorlinks=true,urlcolor=blue,anchorcolor=blue,citecolor=blue,filecolor=blue,linkcolor=blue,menucolor=blue,pagecolor=blue,linktocpage=true]{hyperref}
\usepackage[compress,numbers]{natbib}
\usepackage{braket}
\usepackage{slashed}
\usepackage{physics}
\usepackage{bm}
\usepackage{indentfirst}
\usepackage{appendix}
\usepackage{verbatim}
\usepackage{cancel}
\usepackage[utf8]{inputenc}
\DeclareUnicodeCharacter{2217}{+}
\DeclareUnicodeCharacter{2032}{+}
\usepackage[T1]{fontenc}
\long\def\symbolfootnote[#1]#2{\begingroup%
  \def\thefootnote{\fnsymbol{footnote}}\footnote[#1]{#2}\endgroup}

\renewcommand{\vev}[1]{\langle #1 \rangle}

\newcommand{\mev}{\mathrm{MeV}}
\newcommand{\gev}{\mathrm{GeV}}
\newcommand{\tev}{\mathrm{TeV}}

\input prepictex
\input pictex
\input postpictex
\newdimen\tdim
\tdim=\unitlength

\begin{document}

\begin{titlepage}

  \vspace{0.5cm}
  \begin{center}
    {\Large \bf
      Baryogenesis in a Parity Solution to the Strong CP Problem}
  \end{center}

  \vspace{0.2cm}
  \begin{center}
    {
      Keisuke Harigaya$^{1,2,3,4}$\symbolfootnote[1]{kharigaya@uchicago.edu}
      and Isaac R. Wang$^5$\symbolfootnote[2]{isaac.wang@rutgers.edu}
    }\\
    \vspace{0.6cm}
    \textit{
      \vspace{0.5cm}
      $\,^1$ Department of Physics, University of Chicago, Chicago, IL 60637, USA \\
      \vspace{0.5cm}
      $\,^2$ Enrico Fermi Institute and Kavli Institute for Cosmological Physics, University of Chicago, Chicago, IL 60637, USA \\
      \vspace{0.5cm}
      $\,^3$ Kavli Institute for the Physics and Mathematics of the Universe (WPI),\\
      The University of Tokyo Institutes for Advanced Study,\\
      The University of Tokyo, Kashiwa, Chiba 277-8583, Japan \\
      \vspace{0.5cm}
      $\,^4$ Theoretical Physics Department,
      CERN,
      1211 Geneva 23, Switzerland\\
      \vspace{0.5cm}
      $\,^5$New High Energy Theory Center,
      Department of Physics and Astronomy,\\
      Rutgers University, Piscataway, NJ 08854, USA}\\
  \end{center}

  \vspace{0.4cm}

  \begin{abstract}
    Space-time parity can solve the strong CP problem and introduces a spontaneously broken $SU(2)_R$ gauge symmetry. We investigate the possibility of baryogenesis from a first-order $SU(2)_R$ phase transition similar to electroweak baryogenesis. We consider a model with the minimal Higgs content, for which the strong CP problem is indeed solved without introducing extra symmetry beyond parity. Although the parity symmetry seems to forbid the $SU(2)_R$ anomaly of the $B-L$ symmetry, the structure of the fermion masses can allow for the $SU(2)_R$ sphaleron process to produce non-zero $B-L$ asymmetry of Standard Model particles so that the wash out by the $SU(2)_L$ sphaleron process is avoided. The setup predicts a new hyper-charged fermion whose mass is correlated with the $SU(2)_R$ symmetry breaking scale and hence with the $SU(2)_R$ gauge boson mass, and depending on the origin of CP violation, with an electron electric dipole moment. In a setup where CP violation and the first-order phase transition are assisted by a singlet scalar field, the singlet can be searched for at future colliders.
  \end{abstract}

\end{titlepage}

\vspace{0.2cm}
\noindent

\noindent\makebox[\linewidth]{\rule{\textwidth}{1pt}}
\tableofcontents
\noindent\makebox[\linewidth]{\rule{\textwidth}{1pt}}
\newpage

\section{Introduction}

The CP phases in the quark masses, which explain
the CP violation in the weak interaction,  are expected to also introduce CP violation in the strong interaction~\cite{Bell:1969ts,Adler:1969gk,tHooft:1976rip,tHooft:1976snw}. However, the CP-violating phase in the strong interaction
is known to be smaller than $10^{-10}$ from the non-observation of the neutron electric dipole moment (EDM)~\cite{nEDM:2020crw}. This discrepancy is called the strong CP problem.

The absence of the strong CP violation can be explained by a space-time parity symmetry~\cite{Beg:1978mt,Mohapatra:1978fy}, which predicts the parity partner of the $SU(2)_L$ gauge symmetry called the $SU(2)_R$ gauge symmetry. The $SU(2)_R$ gauge symmetry must be spontaneously broken at an energy scale higher than the electroweak scale to explain the absence of $SU(2)_R$ gauge bosons at the electroweak scale.

A model with an $SU(2)_L$-doublet Higgs $H_L$ and an $SU(2)_R$-doublet Higgs $H_R$ is particularly appealing since the Higgs potential does not contain any physical phases and the strong CP problem is indeed solved without introducing extra symmetries~\cite{Babu:1988mw,Babu:1989rb}; see~\cite{Beg:1978mt,Mohapatra:1978fy,Kuchimanchi:1995rp,Mohapatra:1995xd} for setups with a different Higgs content and additional symmetries. Quantum corrections to the strong CP phase are computed in~\cite{Hall:2018let,deVries:2021pzl} and are found to be sufficiently small.

In this paper, we pursue a possible cosmological role of the $SU(2)_R$ gauge symmetry; production of the baryon asymmetry of the universe. The baryon symmetry has $SU(2)_R$ anomaly and is violated by $SU(2)_R$ sphaleron processes. As in electroweak baryogenesis~\cite{Kuzmin:1985mm}, the baryon asymmetry of the universe may be produced from this baryon number violation, a first-order $SU(2)_R$ phase transition, and some CP violation in the early universe.

There seems to be an apparent obstacle to this idea. The Standard model (SM) $B-L$ symmetry does not have $SU(2)_L$ anomaly, so the parity symmetry seems to require that $B-L$ symmetry also does not have $SU(2)_R$ anomaly. The first-order $SU(2)_R$ phase transition cannot create $B-L$ asymmetry, and baryon asymmetry produced by the $SU(2)_R$ phase transition will be immediately washed out by the $SU(2)_L$ sphaleron process.
This naive expectation assumes that the asymmetry of $SU(2)_R$ charged particles is rapidly transferred into that of $SU(2)_L$ charged particles.
This is indeed the case in the models where the SM Higgs and right-handed fermions are embedded into a bi-fundamental Higgs $\Phi$ and $SU(2)_R$ doublet fermions $\bar{f}$ respectively, and the SM Yukawa couplings come from $f \Phi \bar{f}$, where $f$ are SM $SU(2)_L$-doublet fermions. As we will see, this is not necessarily the case in the model with $H_L$ and $H_R$,
and the washout may be avoided.

Our scenario predicts a hyper-charged fermion with a mass given by $m_{i} v_R/v_L$, where $i$ is $e$, $\mu$, or $\tau$, and $v_R$ and $v_L$ are $SU(2)_R$ and $SU(2)_L$ symmetry breaking scale, respectively.
The mass of the new fermion is then correlated with the masses of $SU(2)_R$ gauge bosons, and with the electron EDM, depending on the source of CP violation.
The case with $i=\tau$ is particularly interesting since $v_R$ predicted from the allowed range of the new fermion mass ($\gtrsim 100$ GeV) overlaps with $v_R$ that is accessible at near future colliders and measurements of the electron EDM.

The first-order $SU(2)_R$ phase transition may be achieved as in electroweak baryogenesis, namely, by a thermal potential from a quartic coupling~\cite{Bochkarev:1987wf,Anderson:1991zb,Espinosa:1993bs, Espinosa:1993yi,Bodeker:1996pc,Carena:1996wj,Espinosa:1996qw,Carena:1997ki,Cline:1998hy}
or an (effective) tree-level trilinear coupling~\cite{Pietroni:1992in,Choi:1993cv,Turok:1991uc,Cline:1995dg,Cline:1996mga,Barger:2008jx} of $H_R$ with a new scalar.
We consider two examples without hierarchy problems beyond that of $H_L$ and $H_R$.
We analyze a model with singlet scalar fields with the minimal coupling to $H_L$ and $H_R$ that was analyzed for electroweak baryogenesis~\cite{Das:2009ue,Harigaya:2022ptp}.
We may also utilize the running of the Higgs quartic coupling; the quartic coupling of $H_L$ becomes small at high energy scales, so if $v_R$ is sufficiently high, the quartic coupling of $H_R$, which is equal to that of $H_L$ evaluated at $v_R$, can be small enough for a first-order $SU(2)_R$ phase transition to occur. If the running is induced only by the SM interaction, $v_R$ is required to be above $10^8$ GeV, but extra interactions of $H_L$ can lower $v_R$.

CP violation can be obtained in various ways. Note that the parity symmetry does not forbid the CP phases of the theory; it only puts relations among them.
As an example, we consider CP violation from a dimension-6 coupling between $H_R$ and the $SU(2)_R$ gauge bosons~\cite{Dine:1990fj} and that from a dimension-5 coupling between a singlet scalar and the $SU(2)_R$ gauge bosons. Those couplings induce non-zero EDMs of SM fermions. That of an electron is detectable by near future experiments if $v_R = O(10)$ TeV, for which the new hyper-charged fermion is within the reach of near-future colliders.

There are several past works on first-order $SU(2)_R$ phase transitions. Ref.~\cite{Brdar:2019fur} investigated $SU(2)_R$ breaking by a triplet scalar and computed the resultant gravitational-wave spectrum. Ref.~\cite{Fujikura:2021abj} considers a model of baryogenesis with extra chiral $SU(2)_R$ charged fermions and the possible embedding of the model into a parity-symmetric theory by further extending the gauge group at UV. See~\cite{Shu:2006mm,Shelton:2010ta,Davoudiasl:2016yfa,Fornal:2017owa,Hall:2019ank} for baryogenesis models from a new non-Abelian gauge symmetry other than $SU(2)_R$.

This paper is organized as follows. In Sec.~\ref{sec:parity}, we describe a parity-symmetric model with $H_L$ and $H_R$. We discuss how the fermion masses, including the neutrino masses, are obtained. In Sec.~\ref{sec:BG}, we present a model of baryogenesis from the $SU(2)_R$ phase transition. Experimental signals are discussed in Sec.~\ref{sec:signal}. Finally, a summary and discussion are given in Sec.~\ref{sec:summary}.

\section{A parity-symmetric model}
\label{sec:parity}
In this section, we describe a parity-symmetric model we study.

\subsection{Gauge symmetry breaking}

The gauge symmetry at UV is $SU(3)_c\times SU(2)_L\times SU(2)_R\times U(1)_{X}$, which is broken down to the SM gauge symmetry by the vacuum expectation value (VEV) of $H_R({\bf 1},{\bf 1},{\bf 2},1/2)$. The SM gauge symmetry is broken down to $SU(3)_c\times U(1)_{\rm EM}$ by the VEV of $H_L({\bf 1},{\bf 2},{\bf 1},-1/2)$.

We impose a parity symmetry $H_R\leftrightarrow H_L^\dag$ that solves the strong CP problem~\cite{Babu:1988mw,Babu:1989rb} as we will see.
The parity-symmetric potential of $H_L$ and $H_R$ is
\begin{align}
  \label{eq:HL HR potential}
  V = \lambda\left(|H_R|^4 + |H_L|^4\right) + \lambda_{LR} |H_L|^2|H_R|^2 - m^2\left(|H_R|^2 + |H_L|^2\right).
\end{align}
The mass of the $SU(2)_R$ gauge boson must be much larger than that of the electroweak gauge boson.
At the tree level, however, there is no vacuum with $v_R > v_L \neq 0 $ for any choice of the parameters of potential. Such a phenomenologically viable vacuum can be obtained by softly breaking the parity symmetry~\cite{Babu:1988mw,Babu:1989rb},
\begin{align}
  \Delta V = -\Delta m^2 \left(|H_R|^2 - |H_L|^2\right), ~~\Delta m^2 > 0,
\end{align}
or by quantum corrections to the Higgs potential~\cite{Hall:2018let}. In this paper, we consider the former option, since the latter option leads to the production of domain walls upon $SU(2)_R$ phase transition. The soft breaking may be understood as spontaneous breaking by a field that couples to $H_L$ and $H_R$.

This theory in general has three tuning; small $m^2 / \Lambda_{\rm UV}^2 \sim v_R^2/ \Lambda_{\rm UV}^2$, small $\Delta m^2 /\Lambda_{\rm UV}^2 \sim v_R^2 /\Lambda_{\rm UV}^2$, and small $(m^2 - \Delta m^2)/m^2 \sim v_L^2/v_R^2$, where $\Lambda_{\rm UV}$ is the UV scale. One may remove the second one by generating $\Delta m^2$ by dynamical transmutation. The total fine-tuning is $v_R^2/ \Lambda_{\rm UV}^2 \times v_L^2/v_R^2 = v_L^2/ \Lambda_{\rm UV}^2 $ and is the same as the SM. This tuning may be explained by anthropic principle~\cite{Agrawal:1997gf,Hall:2014dfa,DAmico:2019hih}. One can also remove the first tuning by embedding the theory into solutions to the EW hierarchy problems with a mass scale $\sim v_R^2$, such as supersymmetric or composite scenarios. The last tuning $v_L^2/v_R^2$, however, cannot be removed.

\subsection{Charged fermion masses}

\begin{table}[tbp]
  \caption{The gauge charges of Higgses and fermions}
  \begin{center}
    \begin{tabular}{|c|c|c|c|c|c|c|c|c|c|c|c|c|} \hline
                & $H_L$          & $H_R$         & $q_i$         & $\bar{q}_i$     & $\ell_i$        & $\bar{\ell}_i$ & $U_i$         & $\bar{U}_i$     & $D_i$          & $\bar{D}_i$     & $E_i$   & $\bar{E}_i$ \\ \hline
      $SU(3)_c$ & {\bf 1}        & {\bf 1}       & {\bf 3}       & ${\bf \bar{3}}$ & {\bf 1}         & {\bf 1}        & {\bf 3}       & ${\bf \bar{3}}$ & {\bf 3}        & ${\bf \bar{3}}$ & {\bf 1} & {\bf 1}     \\
      $SU(2)_L$ & {\bf 2}        & {\bf 1}       & {\bf 2}       & {\bf 1}         & {\bf 2}         & {\bf 1}        & {\bf 1}       & {\bf 1}         & {\bf 1}        & {\bf 1}         & {\bf 1} & {\bf 1}     \\
      $SU(2)_R$ & {\bf 1}        & {\bf 2}       & {\bf 1}       & {\bf 2}         & {\bf 1}         & {\bf 2}        & {\bf 1}       & {\bf 1}         & {\bf 1}        & {\bf 1}         & {\bf 1} & {\bf 1}     \\
      $U(1)_X$  & $-\frac{1}{2}$ & $\frac{1}{2}$ & $\frac{1}{6}$ & $-\frac{1}{6}$  & $- \frac{1}{2}$ & $\frac{1}{2}$  & $\frac{2}{3}$ & $-\frac{2}{3}$  & $-\frac{1}{3}$ & $\frac{1}{3}$   & $-1$    & $1$         \\ \hline
    \end{tabular}
  \end{center}
  \label{tab:charges}
\end{table}%

The quark and lepton masses may be given by the following Yukawa interaction and masses,
\begin{align}
  \label{eq:fermion masses}
  {\cal L} = & x^u_{ij} q_i \bar{U}_j H_L^\dag + \bar{x}^u_{ij} \bar{q}_i U_j H_R^\dag + M_{ij}^u U_i \bar{U}_j  \nonumber \\
  +          & x^d_{ij} q_i \bar{D}_j H_L + \bar{x}^d_{ij} \bar{q}_i D_j H_R + M_{ij}^d D_i \bar{D}_j \nonumber            \\
  +          & x^e_{ij} \ell_i \bar{E}_j H_L + \bar{x}^e_{ij} \bar{\ell}_i E_j H_R + M_{ij}^e E_i \bar{E}_j + {\rm h.c.}
\end{align}
The gauge charges of the fermions are listed in Table~\ref{tab:charges}. Other combinations of fermions are possible as systematically investigated in~\cite{Hall:2018let}, but to be concrete, we focus on this case in this paper.
The parity symmetry restricts the form of $x$, $\bar{x}$, and $M$ as we will see. See Refs.~\cite{Kawasaki:2013apa,Babu:2018vrl,Craig:2020bnv} for the flavor phenomenology of the setup.

In the limit of $M \gg \bar{x} v_R$, we may integrate out the Dirac fermions and obtain dimension-5 operators of the form
\begin{align}
  \label{eq:dim-5 yukawa}
  x\frac{1}{M}\bar{x}^t f \bar{f} H_L^{(\dag)} H_R^{(\dag)},
\end{align}
where $f = q,\ell$.
The SM fermion Yukawa couplings are given by $x \bar{x}v_R/M $. In this limit, the right-handed SM fermions $(\bar{u},\bar{d},\bar{e})$ are dominantly from $SU(2)_R$ doublets $\bar{q}$ and $\bar{\ell}$.
In the opposite limit $M\ll \bar{x} v_R$, heavy fermions obtain masses of $\bar{x} v_R$ and the SM Yukawas are given by $x$. The right-handed SM fermions are dominantly from $SU(2)_R$ singlets $\bar{U}$, $\bar{D}$, and $\bar{E}$.
Whether or not $M \gg \bar{x} v_R$ can depend on the fermion species and generations; we consider such cases in Sec.~\ref{sec:BG} to obtain $B-L$ asymmetry.
From the collider searches on extra quarks~\cite{ATLAS:2015lpr,CMS:2017asf,ATLAS:2018ziw,CMS:2018zkf}, a large Dirac mass term $M \gg xv_R$ is required for the first generation unless $v_R \geq 10^8~\gev$.

\subsection{Parity and strong CP phase}

With the $SU(2)_R$ gauge symmetry, we may impose a space-time parity symmetry.
It acts on gauge fields as
\begin{align}
  G_\mu^a(t,x)  \rightarrow G_\mu^a(t,-x)\times s(\mu),~
  B_\mu(t,x)  \rightarrow B_\mu(t,-x)\times s(\mu),
  \nonumber \\
  W_{L,\mu}^a(t,x)  \rightarrow W_{R,\mu}^a(t,-x)\times s(\mu),~
  W_{R,\mu}^a(t,x)  \rightarrow W_{L,\mu}^a(t,-x)\times s(\mu),
  \nonumber \\
  s(\mu)=\begin{cases}
           1  & \mu = 0     \\
           -1 & \mu = 1,2,3
         \end{cases}
\end{align}
and forbids the CP violating phase in $\theta G\tilde{G}$.
The action on fermions is
\begin{align}
  q(t,x) \rightarrow i \sigma_2 \bar{q}^*(t,-x), ~U(t,x) \rightarrow i \sigma_2 \bar{U}^*(t,-x),~\cdots.
\end{align}
This requires that $\bar{x}_{ij} = x_{ij}^*$ and $M_{ij} = M_{ji}^*$. As a result, the quark mass matrix becomes
\begin{align}
  \begin{pmatrix}
    u_i & U_i
  \end{pmatrix}
  \begin{pmatrix}
    0            & x_{ij} v_L \\
    x_{ji}^* v_R & M_{ij}
  \end{pmatrix}
  \begin{pmatrix}
    \bar{u}_j \\ \bar{U}_j
  \end{pmatrix}.
\end{align}
The determinant of the mass matrix is real, and at the one-loop level, the strong CP phase is not generated from the quark mass~\cite{Babu:1988mw,Babu:1989rb}. Note also that the Higgs VEVs do not have physical phases, since the phases of the Higgses are gauge degrees of freedom. This is an advantage of the gauge symmetry breaking by $H_R$ and $H_L$ in comparison with that by an $SU(2)_R$ triplet and $SU(2)_R\times SU(2)_L$ bi-fundamentals, where the physical phases of Higgs VEVs must be forbidden by extra symmetries~\cite{Beg:1978mt,Mohapatra:1978fy,Kuchimanchi:1995rp,Mohapatra:1995xd}.

Two-loop corrections to the phase of the determinant of the quark mass matrix are estimated in Refs.~\cite{Hall:2018let,deVries:2021pzl} for $M > x v_R$ for all quarks and are found to be below the current upper bound. Refs.~\cite{Babu:1988mw,Babu:1989rb} introduce soft breaking of the parity to the Dirac mass $M_{ij}$. Although the determinant of the quark mass matrix is real at the tree level,  one-loop correction to the phase is generically too large~\cite{deVries:2021pzl}. We assume that the soft breaking in the Dirac mass is suppressed; this is natural given that the Dirac masses are dimension-3 operators while the Higgs masses are dimension-2 operators, and that the Dirac mass may be protected by chiral symmetry.

\subsection{Neutrino masses}
\label{sec:neutrino masses}

The masses of SM and right-handed neutrinos, $\nu$ and $\bar{N}$, can be obtained in several ways.
In one of them discussed in Sec.~\ref{sec:Majorana}, right-handed neutrinos can be dark matter.

\subsubsection{Majorana mass from dimension-five operators}
\label{sec:Majorana}
Neutrino masses can arise from the following two dimension-5 terms,
\begin{align}
  \label{eq:dim-5 Majorana neutrino}
  c^M_{ij} \ell_i\ell_j H_L^{\dag^2} + c^{M*}_{ij} \bar{\ell}_i\bar{\ell}_j H_R^{\dag^2} + {\rm h.c.},
\end{align}
which can be UV-completed by the see-saw mechanism~\cite{Yanagida:1979as,GellMann:1980vs,Minkowski:1977sc,Mohapatra:1979ia}.
The sum of the right-handed neutrino masses is given by
\begin{align}
  \sum_i m_{\nu_i} (\frac{v_R}{v_L})^2 = 3~{\rm keV} \frac{\sum_i m_{\nu_i}}{60~{\rm meV}} \left( \frac{v_R}{40~{\rm TeV}}\right)^2.
\end{align}

For experimentally allowed $v_R > 20$ TeV, the thermal abundance of the right-handed neutrinos exceed the observed dark matter density. This problem can be avoided by entropy production, and the right-handed neutrinos are good dark matter candidates. The required amount of dilution is
\begin{align}
  \label{eq:dilution}
  D = 40 \frac{\sum_i m_{\nu_i}}{60~{\rm meV}} \left( \frac{v_R}{40~{\rm TeV}} \right)^2 \frac{80}{g_s(T_D)},
\end{align}
where $T_D$ is the temperature when the right-handed neutrinos decouple from the thermal bath and $g_s$ is the entropy degree of freedom. This entropy production also dilutes the baryon asymmetry produced at the $SU(2)_R$ phase transition,
and as we will see in Sec.~\ref{sec:scalar extension}, $v_R$ is bounded from above.

The keV-scale right-handed neutrino dark matter is warm and constrained by the observations of the small-scale structure. Observations of Ly-$\alpha$ forests give a constraint $m_N > $ few keV~\cite{Viel:2013fqw,Irsic:2017ixq,Palanque-Delabrouille:2019iyz,Garzilli:2019qki}, which is satisfied if $v_R \gtrsim 40$ TeV.

The right-handed neutrinos can decay into a SM neutrino and a photon. The decay is induced by a one-loop diagram
where $N$ splits into an off-shell $W_R$ and an $SU(2)_R$ charged lepton, they mix with a $W_L$ and an $SU(2)_L$ charged lepton respectively, they annihilate into $\nu$, and a photon is attached to an electromagnetically charged particle inside the loop~\cite{Bezrukov:2009th,Greljo:2018ogz,Dror:2020jzy}.
Here the Dirac mass term of $SU(2)_L\times SU(2)_R$ singlet fermions ($M_{ij}$ in Eq.~\eqref{eq:fermion masses}) is necessary for the $\ell-\bar{\ell}$ and $W_L-W_R$ mixing to exist. If those of the third-generation fermions are suppressed, the constraint derived in Ref.~\cite{Dror:2020jzy} is relaxed. Indeed, as we will see in Sec.~\ref{sec:BG}, the Dirac mass term for the third-generation charged lepton should be suppressed for successful baryogenesis. This weakens the lower bound on $v_R$ shown in~\cite{Dror:2020jzy} by a factor of $(m_\tau/m_\mu)^{1/2}\simeq 4$, which is already enough to avoid the constraint for $m_N =$ few keV and $v_R\sim 40$ TeV. With a moderately suppressed Dirac mass for the third generation up quark, the constraint is further weakened.

\subsubsection{Dirac mass from dimension-five operators}
\label{sec:Dirac}

Dirac neutrino masses may arise from the following dimension-5 operator,
\begin{align}
  \label{eq:dim-5 Dirac neutrino}
  c^D_{ij} \ell_i \bar{\ell}_j H_L^\dag H_R^\dag + {\rm h.c.},
\end{align}
which can be UV-completed, e.g., by
\begin{align}
  \label{eq:UV-completion of Dirac neutrino}
  {\cal L} = x^\nu \ell H_L^\dag \bar{S} + x^{\nu*} \bar{\ell} H_R^\dag S + M^\nu S \bar{S}.
\end{align}
The right-handed neutrinos have the same mass as the SM ones and behave as dark radiation. For experimentally allowed $v_R > 20$ TeV, right-handed neutrinos decouple from the thermal bath before the QCD phase transition, so $\Delta N_{\rm eff} < 0.3$ and is below the current upper bound~\cite{Planck:2018vyg}.

\subsubsection{Mojorana mass by radiative inverse seesaw}
\label{sec:radiative inverse seesaw}

The right-handed neutrinos can be heavy in the following setup,
\begin{align}
  \label{eq:N with chi}
  \chi_i \left( y^\chi_{ij} \ell_j H_L^\dag + y^{\chi*}_{ij} \bar{\ell}_j H_R^\dag \right) + m_{\chi,i} \chi_i^2,
\end{align}
where $\chi_i$ are gauge-singlet fermions. After $SU(2)_R$ and $SU(2)_L$ breaking, only three linear combinations of the right-handed neutrinos and the SM neutrinos obtain Dirac masses paired with $\chi_i$. Because of $v_R \gg v_L$, the heavy ones are mostly the right-handed neutrinos with masses $y^\chi v_R$.
\footnote{Successful baryogenesis may also be achieved via leptogenesis mechanism from the decaying of the neutral component of $\bar{\ell}$ while $v_R$ is required to be higher than $10^{13}$ GeV~\cite{Carrasco-Martinez:2023nit}.}
It can decay into a SM Higgs and a lepton doublet without leaving any cosmological impacts.

In this setup, the naive lepton number $\ell(+1)$ and $\bar{\ell}(-1)$ is violated by the coupling with $\chi$. This may lead to the wash-out of asymmetry produced by the $SU(2)_R$ phase transition. As we will see in the next section, however, this is not necessarily the case. Rather, the violation helps the generation of the $B-L$ asymmetry of SM particles.

The SM neutrinos remain massless. This can be understood by a $U(1)$ symmetry with charges $\chi(-1)$, $\ell(+1)$, and $\ell (+1)$. In the effective theory, after integrating out heavy fields, the neutrino mass, if exists, should be given by a Majorana mass $\nu^2$, but this is forbidden by the $U(1)$ symmetry. Non-zero neutrino masses can be generated by adding a Majorana mass $m_\chi \chi^2$ that explicitly breaks the $U(1)$ symmetry. The neutrino mass is still zero at tree level, and a non-zero neutrino mass arises at one-loop level, as in the radiative inverse-seesaw model~\cite{Dev:2012sg},
\begin{align}
  m_\nu \sim \frac{(y^{\chi})^2}{16 \pi^2} \frac{m_\chi v_L^2}{(y^\chi v_R)^2} = \frac{1}{16 \pi^2} \frac{m_\chi v_L^2}{v_R^2} \sim 0.1~{\rm eV} \frac{m_\chi}{10~{\rm MeV}} \left( \frac{100~{\rm TeV}}{v_R}\right)^2 ,
\end{align}
where we assume $y^\chi v_R > m_W$.

\section{Baryogenesis from $SU(2)_R$ phase transition}
\label{sec:BG}

In this section, we discuss how a first-order $SU(2)_R$ phase transition and $SU(2)_R$ sphaleron processes can produce baryon asymmetry.
The essential idea is the same as electroweak baryogenesis; the phase transition proceeds through the formation of bubbles, which expand and provide the deviation from thermal equilibrium. The $SU(2)_R$ sphaleron process violates the baryon number outside the bubble to create non-zero baryon asymmetry with the aid of some CP violation. The $SU(2)_R$ sphaleron process becomes ineffective inside the bubbles and the baryon asymmetry is frozen.

There seem to be, however, two apparent obstacles to this idea because of the parity symmetry:
\begin{itemize}
  \item $B-L$ does not have $SU(2)_L$ anomaly, so parity symmetry, which exchange $SU(2)_L$ with $SU(2)_R$, seems to require $B-L$ not to have $SU(2)_R$ anomaly. If so, $SU(2)_R$ phase transition can only create $B$ and $L$ asymmetry with $B-L=0$. Since $SU(2)_L$ gauge symmetry is still unbroken after the $SU(2)_R$ phase transition, the produced asymmetries can be washed out by the $SU(2)_L$ sphaleron process~\cite{Harvey:1990qw}.
        
  \item The quartic coupling of the $SU(2)_R$-breaking Higgs and the $SU(2)_R$ gauge coupling would be as large as the SM counterparts, so the strong first-order phase transition does not seem to take place.
\end{itemize}
We discuss how these two obstacles can be avoided in Secs.~\ref{sec:B-L} and \ref{sec:FOPT}, respectively.
Specifically, our solutions to the first obstacle depend on the aforementioned neutrino and charged lepton mass generation mechanism, and use an effective $B-L$ asymmetry coming from the ineffectiveness of some of the interactions and/or $B-L$ violation in the neutrino sector.  

CP violation can be obtained in various ways.
Note that unlike CP symmetry, parity symmetry does not forbid CP phases but only relates the CP phases of parity partners.
To be concrete, we discuss local baryogenesis~\cite{Dine:1990fj} in Sec.~\ref{sec:localBG}.

\subsection{$B-L$ asymmetry}
\label{sec:B-L}
If the lepton charge of $\bar{\ell}$ is $-1$, i.e., the $B-L$ charge of it is $+1$,
the $B-L$ symmetry does not have $SU(2)_R$ anomaly, so the $SU(2)_R$ sphaleron process does not produce $B-L$ asymmetry. This seems to lead to the washout of baryon asymmetry produced during the $SU(2)_R$ phase transition.
However, what lepton charge to be assigned to $\bar{\ell}$ depends on how its asymmetry is transferred into SM particles.
In the following, we discuss a few examples where the washout is avoided because of this ambiguity.

\subsubsection{Right-handed charged leptons from $SU(2)_R$ doublets}
\label{sec:doublet ebar}

We first consider the case where all of charged-lepton Yukawa interactions are obtained from the dimension-5 operator in Eq.~\eqref{eq:dim-5 yukawa}, for which right-handed SM charged leptons $\bar{e}$ are from $SU(2)_R$ doublets $\bar{\ell}$. The asymmetry in $\bar{e}$ can be then transferred into $\ell$ via the Yukawa coupling. If the scattering between right-handed neutrinos $\bar{N}$ and right-handed electrons $\bar{e}$ via the $W_R$ exchange is also effective,  the lepton charge of $\bar{e}$ and $\bar{N}$ should be $-1$, and the asymmetry is washed out. However, if the $W_R$ exchange decouples before the Yukawa interaction becomes efficient, the wash-out can be avoided.

\paragraph{Light right-handed neutrino}
In the neutrino mass model in Secs.~\ref{sec:Majorana} and \ref{sec:Dirac}, the right-handed neutrino masses are negligible in the early universe. The scattering between $\bar{e}$ and $N$ mediated by $W_R$ is suppressed only by the heavy mass of $W_R$ and decouples at a temperature
\begin{align}
  T_{\rm D} \simeq 10^8~{\rm GeV} \left(\frac{v_R}{10^{10}~{\rm GeV}}\right)^{4/3}.
\end{align}
For $v_R>5\times 10^{7}$ GeV, the $W_R$ exchange decouples before the electron Yukawa becomes effective at a temperature $8.5\times 10^4$ GeV~\cite{Bodeker:2019ajh}.
The asymmetry of $\bar{N}_1$ is not communicated to $\ell$, and the total $B-L$ asymmetry of the SM particles is non-zero,
\begin{align}
  \frac{(B-L)_{\rm SM}}{s} = - \frac{n_{\bar{N}_1}}{s} = - \frac{1}{2} \left.\frac{n_{\bar{\ell}_1}}{s} \right|_{T\sim v_R},
\end{align}
and the wash-out is avoided~\cite{Harigaya:2021txz}. For even higher $v_R$, the muon and tau Yukawa interactions are also out of equilibrium when the $W_R$ exchange decouples, so $B-L$ asymmetry becomes larger.

\paragraph{Heavy right-handed neutrino}
In the neutrino mass model in Sec.~\ref{sec:radiative inverse seesaw}, the right-handed neutrinos can be heavy and their abundance is suppressed for $T\ll y^\chi v_R$.
The scattering rate between $\bar{N}_1$ and $\bar{e}_1$ is suppressed not only by the large $W_R$ mass but also by the Boltzmann factor ${\rm exp}(- y ^\chi v_R/T)$,
\begin{align}
  \Gamma \sim \frac{T^{3/2}(y^\chi v_R)^{7/2}}{8\pi v_R^4}\times {\rm exp}\left(- \frac{y ^\chi v_R}{T}\right ).
\end{align}
For $y^\chi = O(1)$, the $W_R$ exchange decouples before the electron Yukawa gets into equilibrium if $v_R \gtrsim 2\times 10^6$ GeV.

Since the asymmetry in $\bar{N}_1$ is suppressed by the Boltzmann factor unlike the setup with light right-handed neutrinos, one may worry that the total $B-L$ asymmetry is also exponentially suppressed; this may not the case because of the $B-L$ violation by $y^\chi$, which can process the asymmetry produced by the $SU(2)_R$ phase transition into non-zero $B-L$ asymmetry. (This is analogous to the conversion of the asymmetry produced by the GUT baryogenesis, where $B-L=0$, into non-zero $B-L$ asymmetry by lepton-number violation~\cite{Campbell:1992jd,Cline:1993vv,Cline:1993bd,Fukugita:2002hu,Domcke:2020quw}.)  The $B-L$ violation generically leads to the complete wash-out of all asymmetries. However, if $y^\chi$ is nearly diagonal in the charged lepton flavor basis,%
\footnote{The neutrino mixing can come from non-diagonal $m_\chi$.}
the following symmetry, which we call $U(1)_{L_1'}$, is approximately preserved before the electron Yukawa gets into thermal equilibrium at the classical level: $\ell_1(+1)$, $\bar{\ell}_1(+1)$, $\chi_1(-1)$. $B/3-L_1'$ is an approximate symmetry without $SU(3)_c\times SU(2)_L$ gauge anomaly, but has $SU(2)_R$ anomaly and can be produced by the $SU(2)_R$ phase transition. Using the method in~\cite{Harvey:1990qw}, one can show that the standard $B-L$ charge ($\bar{e}_1(+1)$, $\ell_1(-1)$, ...) indeed becomes non-zero. $U(1)_{L_1'}$ is explicitly broken by the electron Yukawa,%
\footnote{The symmetry is also explicitly broken by the Majorana mass term $m_\chi$, but it is still small enough and the wash-out by this explicit breaking is ineffective.}
but by the time it becomes effective, $\bar{e}_1$ no longer communicates its charge with $\bar{N}_1$ via the $W_R$ exchange and the $B-L$ violation by $y^\chi$ is ineffective. Because of the standard $B-L$ charge conservation, the baryon asymmetry remains non-zero.

\subsubsection{A right-handed charged lepton from an $SU(2)_R$ singlet}
\label{eq:singlet ebar}

The scenarios in Sec.~\ref{sec:doublet ebar} require large $v_R$ and there are no light enough particles to be produced at near future colliders. Successful scenarios with lower $v_R$ exist if the  charged lepton masses have the following structure,
\begin{align}
  \label{eq:no Dirac mass}
  {\cal L} = & x^e_{ij} \ell_i \bar{E}_j H_L + x^{e*}_{ij} \bar{\ell}_i E_j H_R + M_{ij}^e E_i \bar{E}_j~~(i,j =1,2) \nonumber \\
             & + y_\tau \ell_3 \bar{E}_3 H_L + y_\tau^* \bar{\ell}_3 E_3 H_R  + {\rm h.c.},
\end{align}
which may be ensured by an approximate $U(1)$ symmetry with charges $\ell_3(1)$, $\bar{\ell}_3(1)$, $E_3(-1)$, and $\bar{E}_3(-1)$. For this structure, the lepton charge of $\bar{\ell}_3$ and $E_3$ are not necessarily $-1$ and $+1$ and the wash-out may be avoided as described below. We call the two components in $\bar{\ell}_3$ as $\bar{N}_3$ and $\bar{\tau}'$, and $E_3$ as $\tau'$. Note that there is no $M_{3i}^e$ or $M_{i3}^e$ in Eq.~\eqref{eq:no Dirac mass}.

Because of the parity symmetry, $\tau'$ mass is predicted to be $m_\tau v_R/v_L$. The collider signature of $\tau'$ is discussed in Sec.~\ref{sec:signal}. 
The mechanisms we discuss below works also for the case where some of the first two generations have the structure of a vanishing Dirac mass, $M^e_{1i}=0$ or $M^e_{2i}=0$, for which even lighter new fermions with masses $y_e v_R$ or $y_\mu v_R$ are predicted. To be conservative, we assume that only $\tau'$ is light.

\paragraph{Light right-handed neutrino}
For the neutrino mass models in Secs.~\ref{sec:Majorana} and \ref{sec:Dirac},
the interaction of $\bar{\ell}$ through the neutrino mass operators is ineffective because of the smallness of the neutrino mass.
With the structure of charged leptons in Eq.~\eqref{eq:no Dirac mass}, among the asymmetry of $\bar{\ell}_{1,2,3}$ created by the $SU(2)_R$ sphaleron process, only that of $\bar{\ell}_{1,2}$ is transferred into $SU(2)_L$ charged particles.
Therefore, the total $B-L$ asymmetry of the SM particles is non-zero,
\begin{align}
  \frac{(B-L)_{\rm SM}}{s} = -\left. \frac{n_{\bar{\ell}_3}}{s}\right|_{T\sim v_R}.
\end{align}
The same way of obtaining non-zero baryon asymmetry is also employed in a model of baryogenesis from axion rotation, called axiogenesis~\cite{Co:2019wyp}, in~\cite{Harigaya:2021txz}.

In this scenario, it is crucial that the transfer of the asymmetries in $\bar{\ell}_3$ and $E_3$ into other leptons is negligible, since otherwise the wash-out of the asymmetry occurs.  Let us derive the bound on the parameters of the theory, taking the mass term $M_3 \bar{E}_3 E_3$ as an example.
Since the scattering rate via $M_3$ is exponentially suppressed at $T < m_{\tau'}$, the ratio between the scattering rate and the Hubble expansion rate is maximized at $T\sim m_{\tau'}$.
A care must be taken in the choice of basis. Instead of choosing the basis where the Yukawa interaction $y_\tau$ is diagonal, we may rotate $(\bar{E}_3, \bar{\tau}')$ to remove $M_3$ and define the asymmetry in this basis.
We should take the basis with a smaller transfer late~\cite{Davidson:1996cc,Davidson:1997mc}, since it is enough to have one basis in which the charges are separated with each other. It turns out that we shall use the basis with a diagonal Dirac mass term when $y_\tau T < m_{\tau'}$, which is indeed the case after $SU(2)_R$ phase transition since $m_{\tau'} = y_\tau v_R$.
This can be also understood from the thermal mass given by the Yukawa interaction $y_\tau T$ being smaller than $m_\tau'$, so that the Hamiltonian of quasi-particles on the thermal background is closer to a diagonal one in the basis where the Dirac mass is diagonalized.
In this basis, the charge transfer is induced by a Yukawa interaction $y_\tau M_3 / m_{\tau'}$.
We require that
\begin{align}
  \label{eq:upM3}
  \alpha_2 y_\tau^2 \frac{M_3^2}{m_{\tau'}^2} T < H(T=m_{\tau'}) \longrightarrow M_3 < 20~{\rm MeV} \left( \frac{v_R}{100~{\rm TeV}}\right)^{3/2}.
\end{align}
Even if this condition is violated, the washout is avoided if the scattering between $\bar{\tau}'$ and $\bar{N}_3$ via the $W_R$ exchange are ineffective when the scattering by $M_3$ becomes effective. However, the $W_R$ exchange is indeed effective at $T\sim m_{\tau'}$, so the upper bound in Eq.~\eqref{eq:upM3} is applicable.

\paragraph{Heavy right-handed neutrino}

We next consider the neutrino mass model in Sec.~\ref{sec:radiative inverse seesaw}.
If the coupling $y^{\chi}_{ij}$ is not nearly diagonal in the charged lepton mass eigenbasis, all possible lepton symmetries are violated, and the asymmetry produced by the $SU(2)_R$ phase transition is washed out.

On the other hand, if $y^{\chi}_{ij}$ is nearly diagonal, although the first and second-generation lepton symmetry is violated, the third-generation leptons preserve the following symmetry which we call $U(1)_{L_3'}$; $\ell_3(+1)$, $\bar{E}_3 (-1)$, $\bar{\ell}_3(+1)$, $E_3(-1)$, and $\chi_3(-1)$.
With this charge assignment, $B/3-L_3'$ has $SU(2)_R$ anomaly and hence can be produced by the $SU(2)_R$ phase transition. As the temperature drops below the right-handed neutrino mass $y^\chi v_R$ and the $\tau'$ mass, the third-generation lepton asymmetry is stored dominantly in $\bar{E}_3$ and $\ell_3$, namely, the tau and tau neutrino. Because of the violation of the standard $B-L$ symmetry by $y^\chi$, the total $B-L$ asymmetry of the SM particles becomes non-zero.

$L_3'$-breaking parameters should be sufficiently small to avoid washout. Let us again take $M_3$ as an example. Note that the washout is avoided even if the scattering or decay involving $M_3$ becomes effective at a low temperature, as long as the scattering between $\tau'$ and $N_3$ has already decoupled by that time. This is because the following standard lepton symmetry is preserved: $\ell_3(+1)$, $\bar{E}_3(-1)$ $\bar{\tau}'(-1)$, $E_3(+1)$. One can show that this standard $B-L$ charge indeed becomes non-zero. At $T\ll y^\chi v_R$, the scattering rate between $\bar{\tau}'$ and $\bar{N}_3$ is suppressed not only by the large $W_R$ mass but also by the Boltzmann factor ${\rm exp}(- y ^\chi v_R/T)$,
\begin{align}
  \Gamma \sim \frac{T^{3/2}(y^\chi v_R)^{7/2}}{8\pi v_R^4}\times {\rm exp}\left(- \frac{y ^\chi v_R}{T}\right ).
\end{align}
For example, for $v_R=20-100$ TeV and $y^\chi= 1$, the scattering decouples at $T_d = 0.7- 4$ TeV, which is above $m_{\tau'}$.
The upper bound on $M_3$ is given by
\begin{align}
  \label{eq:M3 upper bound}
  \alpha_2y_\tau^2 \frac{M_3^2}{m_{\tau^\prime}^2}T \lesssim H(T=T_d) \longrightarrow M_3 < 40 ~{\rm MeV} \left( \frac{T_d}{4~{\rm TeV}}\right)^{1/2} \left(\frac{v_R}{100~\tev} \right).
\end{align}

\subsection{Strong first-order phase transition}
\label{sec:FOPT}

The SM Higgs coupling is about $0.13$ around the weak scale. If the quartic coupling of $H_R$ is of this order and $H_R$ only has the Yukawa couplings in Eqs.~\eqref{eq:HL HR potential} and~\eqref{eq:fermion masses} and gauge interactions, strong first-order phase transition cannot occur. Indeed, as in the SM, for $g_R= g_L\simeq 0.65$, the $SU(2)_R$ Higgs quartic coupling must be smaller than $0.017$~\cite{Kajantie:1995kf,Kajantie:1996mn,Gurtler:1997hr,Rummukainen:1998as,Csikor:1998eu,Jansen:1995yg} for a strong first-order phase transition to occur for vanishing fermion Yukawa contributions, and with heavy top mass the phase transition strength should be further suppressed because of the positive correction to the quartic coupling~\cite{Dine:1992wr, Arnold:1992rz}; see also Appendix~\ref{sec:potential}.

In the rest of this subsection, we discuss how the $SU(2)_R$ phase transition can be of strong first order.
In the models we discuss, no hierarchy problems beyond that of $H_L$ and $H_R$ are introduced.
The effective thermal potential is computed up to one-loop level, as described in Appendix~\ref{sec:potential}. For marginally SFOPT, high-temperature expansion shows a great agreement with the full form. Since we are aiming at finding the boundary of the parameter space to achieve SFOPT, we use the high-temperature expansion throughout this subsection to simplify the computation.

\subsubsection{Running quartic coupling}
\label{sec:running}
With the hierarchy $v_L \ll v_R$, it is possible to utilize the running of the quartic coupling.
The SM Higgs quartic coupling becomes smaller at high energy scales because of the quantum correction from Yukawa couplings. If there are no extra particles contributing to the $\beta$-function of the quartic coupling, the quartic coupling $\lambda_R$ becomes smaller than 0.015 at around $10^8~\gev$. We computed the effective potential for $H_R$ at different energy scales with corresponding running couplings, and found that a SFOPT can be achieved for $v_R > 2 \times 10^8~\gev$. Here and hereafter, we assume that the quartic coupling $|H_L|^2 |H_R|^2$ is negligible. If not, the threshold correction to $|H_L|^4$ at the scale $v_R$ via the $H_R$ exchange makes the SM Higgs quartic coupling $|H_L|^4$ smaller than $|H_R|^4$ and the lower bound on $v_R$ becomes stronger.

If $H_L$ has $O(1)$ Yukawa couplings to new fermions, the running becomes faster, and a strong first-order phase transition can occur for much smaller $v_R$.
As an example, we add extra fermions
\begin{align}
  \label{eq:new vectorlike pair}
  \psi = ({\bf 1},{\bf 2},{\bf 1},-\frac{1}{2}),~\bar{\psi} = ({\bf 1},{\bf 2},{\bf 1},\frac{1}{2}),~ \Psi = ({\bf 1},{\bf 1},{\bf 1},0),
\end{align}
with Yukawa couplings $y_\psi H_L^{\dagger}\psi \Psi$ and $\bar{y}_\psi H_L\bar{\psi}\Psi$ and mass terms $m_\psi \psi \bar{\psi} $ and $m_\Psi \Psi^2$. The corresponding $SU(2)_R$ charged partners and their interactions are also introduced. This extra Yukawa coupling quickly makes the quartic coupling small enough to achieve SFOPT. The required Yukawa coupling $y_{\psi}$ to obtain a small enough $\lambda_R$ depends on the masses of these extra fermions. Their lightness does not introduce extra hierarchy problems because of the protection of fermion masses by chiral symmetry.

\begin{figure}[t]
  \centering
  \includegraphics[width=0.6\linewidth]{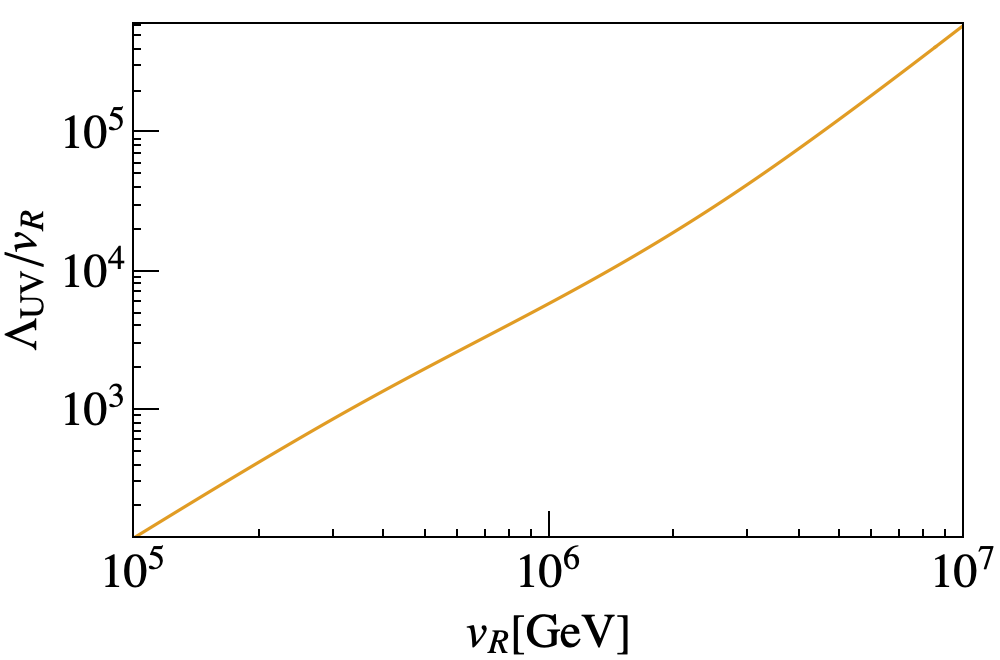}
  \caption{Required scale of new physics $\Lambda_{\rm UV}$ to avoid the instability of the Higgs potential for extra leptons with a mass of 1 TeV. Here the extra Yukawa couplings are chosen to be the minimal value to achieve the SFOPT.}
  \label{fig:UV}
\end{figure}

On the other hand, a large extra Yukawa coupling may introduce the problem of the instability of the potential. With a large extra Yukawa coupling, the quartic coupling at a large Higgs field value quickly becomes negative. The tunneling rate from the metastable point $\vev{H_R}= v_R$ to infinity can be computed via the pseudo-bounce action method~\cite{Espinosa:2019hbm}. We use the zero-temperature effective potential $V_{T=0}$ in Appendix~\ref{sec:potential} to compute the tunneling action.
For an escape point with a large field value, the pseudo-bounce action becomes smaller, making the tunneling more likely to happen. A large enough escape point thus always leads to instability. The value of the escape point that gives a small enough bounce action is regarded as a UV scale $\Lambda_{\rm UV}$ below which the modification of the Higgs potential by new physics is required.
In Fig.~\ref{fig:UV}, we show $\Lambda_{\rm UV}/v_R$ as a function of $v_R$, with the extra Yukawa coupling chosen to be the minimal value achieving SFOPT. The extra fermion assisting the running is assumed to have a mass of $1$ TeV.

We cannot argue that SFOPT can be achieved for $v_R < 10^5$ GeV.
Even for the TeV-scale extra leptons,
the required Yukawa coupling $y_\psi$ is order one, which is close to that of the top quark at the EW scale. For such a large Yukawa coupling, similar to the SM EW phase transition analyzed in Ref~\cite{Dine:1992wr}, the result of 1-loop perturbative computation highly depends on approximation methods and renormalization schemes/scales. Some of them deviate from the result of lattice simulation significantly.
More advanced computation techniques such as dimension reduction, and computation up to higher-order loops are required in such a case~\cite{Croon:2020cgk,Schicho:2021gca,Niemi:2021qvp,Schicho:2022wty}, which is beyond the scope of this paper. For a discussion of the theoretical uncertainties of the computation, see Appendix~\ref{sec:potential}.

We comment on the uncertainty from the top quark mass, to which the running is sensitive.
We used the central value $m_t = 172.69~\gev$~\cite{Workman:2022ynf}. If the top quark mass shifts by  $1$ GeV, the minimal $v_R$ achieving SFOPT in the minimal model changes by a factor of $10$. The prediction for $\Lambda_{\rm UV}/v_R$ in Fig.~\ref{fig:UV}, however, does not change, since the shift of the top quark mass can be absorbed into the shift of the free parameters $y_{\psi}$ and $\bar{y}_{\psi}$.

\subsubsection{Extra scalars}
\label{sec:scalar extension}

A strong first-order phase transition may be also achieved via extra scalar fields that couple to Higgses, which have been intensively investigated in the literature in the context of electroweak phase transition.
Those models can be categorized into two types: the couplings may enhance the cubic term (see Appendix~\ref{sec:potential}) via loop effects, or modify the (effective) tree-level quartic coupling directly via mixing with the Higgs. The former includes, for example, the MSSM~\cite{Espinosa:1993yi,Bodeker:1996pc,Carena:1996wj,Espinosa:1996qw,Carena:1997ki,Cline:1998hy}, a real $Z_2$-even singlet scalar~\cite{Anderson:1991zb,Espinosa:1993bs}, and so forth.
The latter includes the 2HDM model~\cite{Turok:1991uc,Cline:1995dg,Cline:1996mga}, a real $Z_2$-odd singlet scalar~\cite{Choi:1993cv}, the NMSSM~\cite{Pietroni:1992in}, a complex singlet scalar~\cite{Barger:2008jx}, and so forth.
For a review, see~\cite{Morrissey:2012db}.
Here we analyze an example of real singlet scalars investigated in~\cite{Das:2009ue,Harigaya:2022ptp}, where the singlet scalars are naturally light and no extra hierarchy problem is introduced.

We introduce two extra scalar singlets  $S_L$ and $S_R$. (One singlet case does not work as we explain later.) The tree-level potential is
\begin{align}
  \label{eq:scalar extension potential}
  V_0 = & -\frac{1}{2}\mu_{H_L}^2h_L^2 - \frac{1}{2}\mu_{H_R}^2 h_R^2 + \frac{1}{4}\lambda (h_L^4 + h_R^4) + \frac{1}{4}\lambda_{\rm LR} h_L^2 h_R^2 \nonumber \\
        & + \frac{1}{2} \mu_S^2 (S_L^2 + S_R^2) + \frac{1}{2}A S_L (h_L^2 - 2 v_L^2) + \frac{1}{2}A S_R (h_R^2 - 2v_R^2)\nonumber                              \\
        & + \frac{1}{2}A' S_L (h_R^2 - 2 v_R^2) + \frac{1}{2}A' S_R (h_L^2 - 2v_L^2).
\end{align}
$S_L$ and $S_R$ enjoy a shift symmetry $S_{L,R}\rightarrow S_{L,R}+ \delta_{L,R}$ that is softly broken by the mass terms and the trilinear couplings. As long as $A,A' < \mu_S$, the lightness of $S_{L,R}$ is natural.

Let us for now take $A'$ to be zero and discuss the effect of non-zero $A'$ later.
For given field values of $h_L$ and $h_R$, the potential is minimized at
\begin{align}
  \label{eq:S vev}
  \langle S_{{L,R}} \rangle = \frac{A}{2\mu_{S}^{2}} (h_{L,R}^2 - 2 v_{L,R}^2).
\end{align}
The potential along this trajectory is
\begin{align}
  \label{eq:1d scalar potential}
  V_0(h_L,h_R,\langle S_L \rangle, \langle S_R \rangle) = - \frac{1}{2}\mu_{H_L}^2 h_L^2 - \frac{1}{2} \mu_{H_R}^2 h_R^2 + \frac{1}{4} (\lambda - \frac{A^2}{2 \mu_S^2}) (h_L^4 + h_R^4) + \frac{1}{4} \lambda_{\rm LR}h_L^2 h_R^2.
\end{align}
One can see that the effective quartic coupling $\lambda_{\rm eff}$ receives a tree-level modification, $\lambda_{\rm eff} = \lambda(v_R) - A^2/(2\mu_S^2)$, and the phase transition strength is enhanced.

\begin{figure}[!t]
  \centering
  \begin{minipage}[t]{0.496\linewidth}
    \includegraphics[width=3.2in]{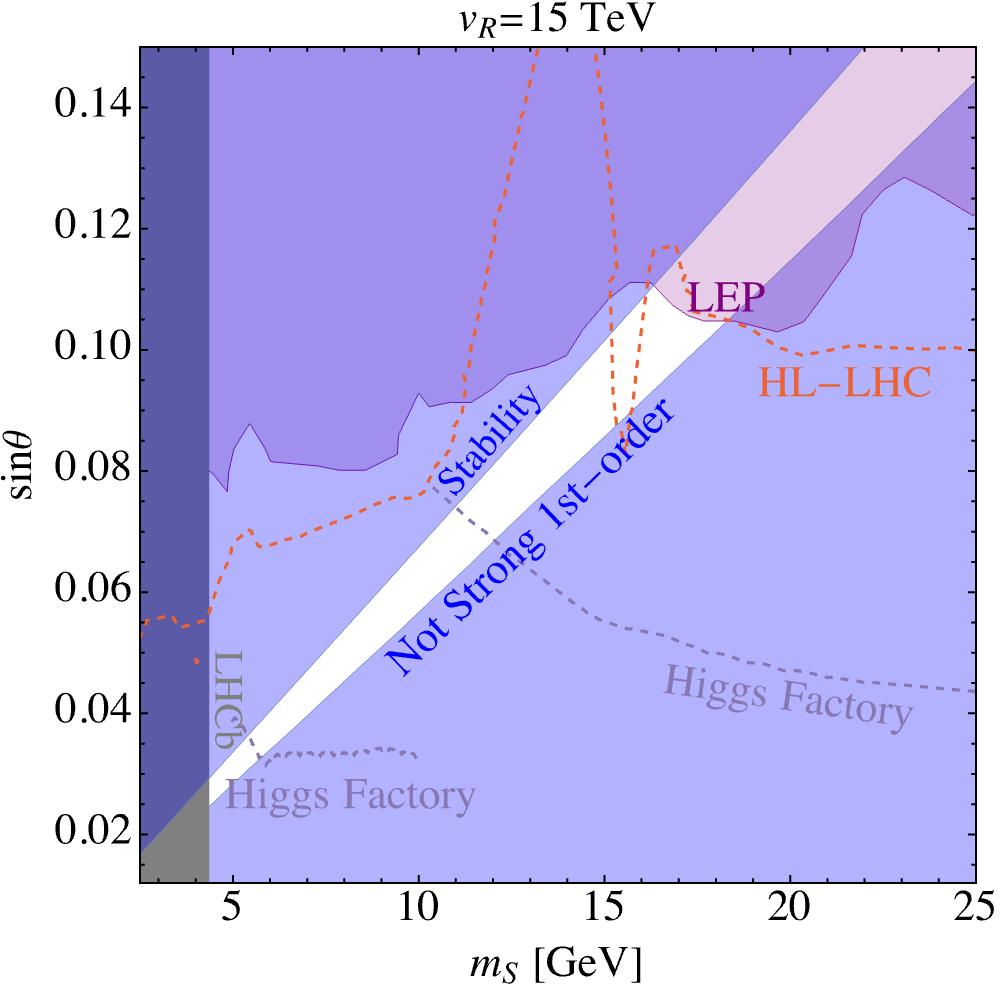}
  \end{minipage}
  \begin{minipage}[t]{0.496\linewidth}
    \includegraphics[width=3.2in]{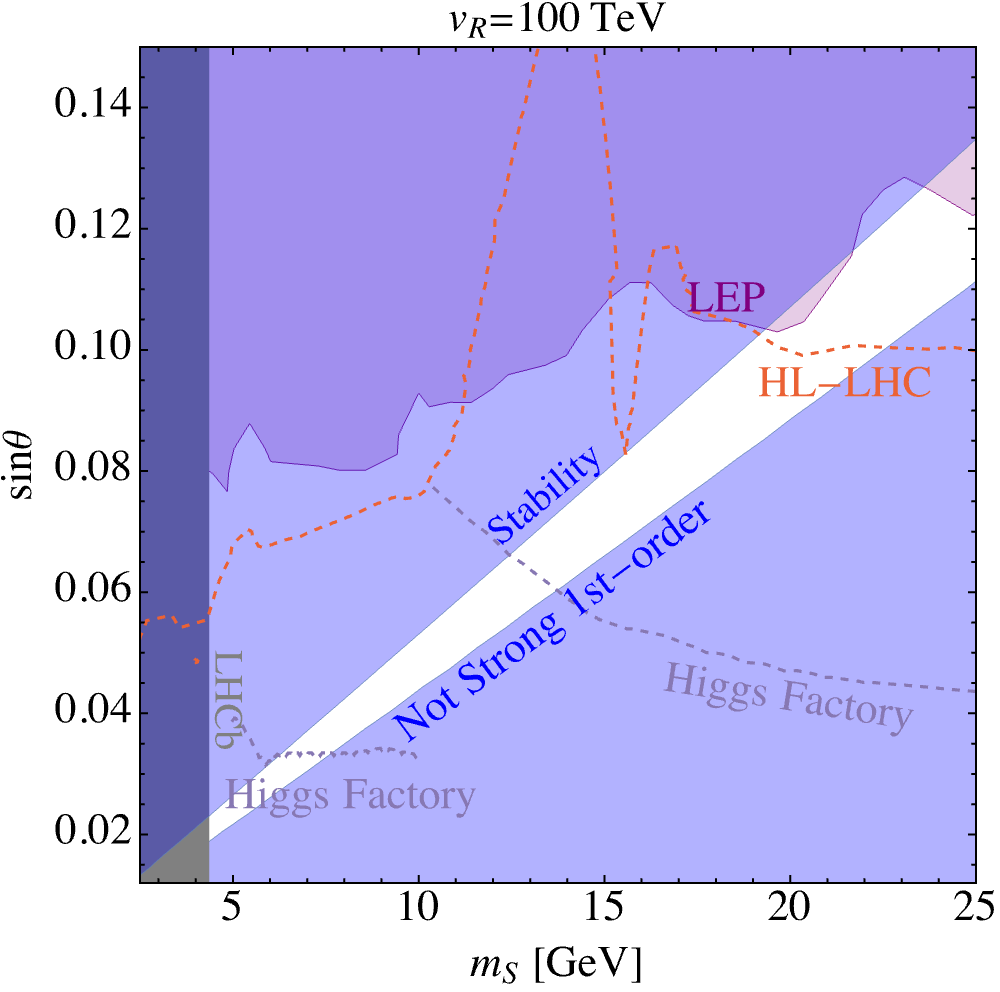}
  \end{minipage}
  \caption{Parameter space for the scalar extension in Eq.~\eqref{eq:scalar extension potential}. Blue-shaded region: excluded by the SFOPT condition and stability requirement. Magenta-shaded region: LEP constraint from direct production. Orange-shaded region: LHC constraint from  Higgs exotic decay. Gray shaded region: LHCb constraint from a rare B-meson decay.
    White space: allowed parameter space.
    Orange dashed line: HL-LHC prospect for extra scalar search via Higgs exotic decay. Purple dashed line: Higgs factory prospects for extra scalar search via Higgs exotic decay.}
  \label{fig:scalar}
\end{figure}

The lower bound on $A$ to achieve a SFOPT is translated into a lower bound on $S_L-h_L$ mixing, which is shown by the lower blue-shaded regions in Fig.~\ref{fig:scalar}. We take into account the running of $A$ and $\lambda$ from the EW scale to $v_R$.
We take $\lambda_{\rm LR}$ to be zero; non-zero $\lambda_{\rm LR}$ gives a tree-level threshold correction to the quartic coupling of $H_L$ via the $H_R$ exchange and the required $A$ becomes larger. $\lambda_{\rm LR}$ induced by $U(1)_{X}$ interaction is negligible.
Since the quartic coupling $\lambda$ is smaller at $v_R$ than at the EW scale, the required magnitude of $A$ to achieve the SFOPT becomes smaller and thus mixing angle is smaller than that in Ref.~\cite{Das:2009ue,Harigaya:2022ptp}. In the upper blue-shaded region, $\lambda - A^2/(2\mu_S^2)$ at $v_R$ is negative, and the potential is unstable.
Here we again assume $m_t = 172.69~\gev$~\cite{Workman:2022ynf}. We found that the prediction for the mixing angle shifts by $5\%$ for the shift of the top quark mass by $1$ GeV.
We discuss the uncertainties in the computation of the electroweak-like phase transition in Appendix~\ref{sec:potential}. The result agrees with the conclusion of Ref.~\cite{Quiros:1999jp,Arnold:1992rz} that 2-loop computation  gives a stronger phase transition. Thus, our prediction in Fig~\ref{fig:scalar}, which is based on the 1-loop potential with high-temperature expansion, is regarded as a conservative one.

Here it is crucial that the mass of $S$ is below the EW scale and cannot be integrated out. Otherwise, the effective coupling of $H_L$ corrected by $A$ is no longer an effective one. Rather, it is the actual Higgs coupling in the effective theory after integrating out $S$ that determines the SM Higgs mass. As a result, the quartic coupling at the energy scale above the $S$ mass becomes larger and the $SU(2)_R$ phase transition cannot be a SFOPT unless $v_R$ is above $10^8$ GeV.

The model is further constrained by direct searches for the singlet scalars.
The mixing between $S_R$ and the SM Higgs is suppressed by $\lambda_{\rm LR}$, so we focus on $S_L$ that mixes with the SM Higgs and can be probed in various ways.
$S_L$ heavier than a few GeV can be probed in collider experiments by direct production of $S_L$.
A search up to $100$ GeV was performed assuming that $S_L$ decays into SM fermions, with $S_L\rightarrow b \bar{b}$ providing the most stringent bound~\cite{LEPWorkingGroupforHiggsbosonsearches:2003ing}.
A search independent of the decay mode of $S_L$ via $e^+e^- \rightarrow Z S_L$ is also performed~\cite{L3:1996ome, OPAL:2002ifx}. (This search is also applicable even if $S_L$ dominantly decays into a dark sector.)
We compare all the bounds mentioned above and choose the most stringent one for each mass, leading to the magenta-shaded region of Fig.~\ref{fig:scalar}.

This scalar can be also searched via an extra decay channel $h\rightarrow S_L S_L$ of the SM Higgs. This decay channel was searched at LHC for various final states. Ref.~\cite{Carena:2022yvx} summarizes all current searches and derives a combined exclusion curve, using a proper branching ratio for $S_L$ decaying into SM fermions.
The prospect of future search is discussed in Sec.~\ref{sec:scalar future}.
In Fig.~\ref{fig:scalar}, though the current exclusion curve is outside the plotted range, we show the future projection curve with dashed lines.

$S_L$ with a mass below  a few GeV can be probed by rare meson decay (see~\cite{Beacham:2019nyx} for review, and see~\cite{Goudzovski:2022vbt} for recent updates.)
LHCb performed a search with $B^+ \rightarrow K^+S_L(\mu^+\mu^-)$~\cite{LHCb:2016awg} and $B^0 \rightarrow K^0S_L(\mu^+\mu^-)$~\cite{LHCb:2015nkv} for $200~\mev \leq m_S \leq 4~\gev$.
The scalar mass at the MeV scale is not excluded by the direct searches for $S_L$, but is excluded by the cosmological problem of $S_R$ as explained later.

Now let us discuss the effect of $A^\prime$. Without loss of generality we take $A^\prime < A$, otherwise we can rename $S_{L,R}$ as $S_{R,L}$. The most important effect of this coupling is to give a tree-level threshold correction to the mass of $S_L$ via $h_R$ exchange;
\begin{align}
  \label{eq:threshold correction}
  \mu_S^2 & \rightarrow \mu_{S_L}^2 = \mu_S^2 - \frac{A^{\prime 2}}{2\lambda(v_R)}\nonumber                                \\
  \mu_S^2 & \rightarrow \mu_{S_R}^2 = \mu_S^2 - \frac{A^2}{2\lambda(v_R)} = \mu_S^2 \frac{\lambda_{\rm eff}}{\lambda(v_R)}
\end{align}
That means that for a given $\mu_{S_L}^2$, $\mu_S^2$ is larger by $A^\prime/(2\lambda(v_R))$.
To achieve SFOPT, a larger $A$ is required to enhance the tree-level correction to $\lambda_{\rm eff}$, i.e., $-A^2/(2\mu_S^2)$. The allowed parameter space in $(m_S,\theta)$ for SFOPT is shifted above. For example, if $A^\prime$ is not
smaller than $A$ by a factor of 3, the allowed parameter space for SFOPT is excluded by collider experiments. This is also the reason why the model with a single $S$ and a coupling $AS(h_L^2 + h_R^2)$ does not work.

The MeV-scale mass is probed by rare kaon decay~\cite{NA62:2020pwi,NA62:2021zjw} and negative $\Delta N_{\rm eff}$~\cite{Planck:2018vyg,Ibe:2021fed}.
Large mixing with $h_L$ is excluded by the rare kaon decay because of the large branching fraction of $K \rightarrow \pi S $, while small mixing is excluded by the decay of $S$ into electrons after neutrinos decouple, which makes neutrinos relatively cooler.
These constraints on $S_L$ are satisfied for $m_{S_L} < 20$ MeV, but not once the cosmology of $S_R$ is taken into account.
$S_R$ couples with $H_R$ and is thermalized when $T\sim v_R$.
If $A'$ is small, the mixing of $S_R$ with $h_L$ is small, so $S_R$ decouples from the bath when it is relativistic and decays much after neutrinos decouple and $S_R$ becomes non-relativistic. This leads to a too large negative $\Delta N_{\rm eff}$.  If $A^\prime$ is of the same order as $A$, $S_R$ mixes with $h_L$, and the $\Delta N_{\rm eff}$ constraint can be evaded. However, the threshold correction to $\mu^2_{S_L}$ discussed above shifts the allowed parameter space for SFOPT above, and the constraint from the rare kaon decay excludes it. Combining all these requirements, we found no allowed parameter.

\subsection{Example of CPV: Local baryogenesis}
\label{sec:localBG}

In this paper, we consider local baryogenesis for its simplicity and effectiveness, following the proposal in~\cite{Dine:1990fj} for the electroweak phase transition.
We first review the local baryogenesis in a general way for the thick-wall regime where the wall is thicker than the mean free-path of particles $\sim (\alpha_R T)^{-1}$.
Indeed, for the models using the running quartic or the extra singlet scalar, we find that the wall thickness is around $100/T$.
We then discuss two specific models.

If $H_R$ couples to heavy $SU(2)_R$ charged fermions directly or indirectly and CP is violated by the coupling or mass of the fermions, the $SU(2)_R$ theta term may depend on $H_R$,
\begin{align}
  \label{eq:general CP-violation}
  \frac{g^2}{32\pi^2} \theta (H_R) W_R\tilde{W}_R.
\end{align}
We follow the method in~\cite{Dine:1990fj} to calculate the baryon asymmetry. 
During the FOPT by bubble nucleation, spacial points that are initially outside an expanding bubble are later swept by the bubble wall and get inside the bubble.
During this transition, while the spacial points are on the bubble wall, sphaleron processes produce asymmetry of fermions with $SU(2)_R$ charges via quantum anomaly, and the processes become ineffective inside the bubble. 
In the thick-wall regime, the $H_R$ field value changes gradually across the bubble wall and the asymmetry of particles may be computed by thermodynamics and small deviation from the local thermal equilibrium.
The change of the field value of $H_R$ across the bubble wall gives a time-dependent $\theta$ term and hence a bias between the $SU(2)_R$ sphaleron transition increasing asymmetry and that decreasing asymmetry. The imbalance between the two processes produces asymmetry of fermions.  (The produced asymmetry gives an opposite bias, but it is negligible unless the asymmetry reaches an equilibrium value.)

As discussed in Sec.~\ref{sec:B-L}, the final baryon number is determined by the lepton number carried by the new charged lepton $\bar{\ell}_3$ and $E_3$.
The $SU(2)_R$ sphaleron process produces the asymmetry of $\bar{\ell}_3$ as described above,
\begin{align}
  \dot{n}_{\bar{\ell}_3} = \frac{\Gamma_{\rm Rs}}{T^3} \dot{\theta} T^2,
\end{align}
where $\Gamma_{\rm Rs}$ is the $SU(2)_R$ sphaleron transition rate.
This equation can be solved by performing an integration over time on the right-hand side along the wall profile.
When the Higgs field value is sufficiently small, the sphaleron transition rate is given by~\cite{DOnofrio:2014rug}
\begin{align}
  \label{eq:sphaleron rate}
  \Gamma_{\rm Rs} = \kappa \alpha_R^5 T^4,
\end{align}
where $\kappa \simeq 20$.
The sphaleron rate receives an exponential suppression when the field value of $H_R$ becomes large enough. By the time this occurs, $\theta (H_R)$ changes by $\delta \theta$. Then the baryon asymmetry produced by the $SU(2)_R$ phase transition normalized by the entropy density is given by
\begin{align}
  Y_B = \frac{n_B}{s} = \frac{28}{79} \frac{1}{s} \kappa \alpha_R^5 T^3 \delta \theta = 8.6\times 10^{-11} \frac{\delta \theta}{0.02}.
\end{align}
One can see that the observed baryon asymmetry can be obtained even with $\delta \theta \ll 1$.

In the neutrino mass model in Sec.~\ref{sec:Majorana}, right-handed neutrinos can be dark matter if entropy production given by Eq.~\eqref{eq:dilution} occurs. The entropy production also dilutes baryon asymmetry, so larger $\delta \theta$ is required.
This gives an upper bound on the sum of right-handed neutrino mass,
\begin{align}
  \sum_i m_{N_i} < 10~{\rm keV} \times \frac{\delta \theta_{\rm max}}{\pi}.
\end{align}
Here we take the maximal $\delta \theta$, $\delta \theta_{\rm max}$, to be $\pi$. If there are multiple heavy $SU(2)_R$ charged fermions in the UV completion of the local operator in Eq.~\eqref{eq:general CP-violation}, $\delta \theta_{\rm max}$ can be larger.
Unless the number of those heavy fermions is large, the warmness of dark matter can be observed by future observations of 21cm lines~\cite{Munoz:2019hjh}.
The upper bound on the masses of right-handed neutrinos also gives an upper bound on $v_R$,
\begin{align}
  v_R \leq 70~\tev \left(\frac{60 \mathrm{meV}}{\sum_i m_{\nu_i}}\right)^{1/2} \left(\frac{g_s(T_D)}{80}\right)^{1/2} \left(\frac{\delta \theta_{\rm max}}{\pi}\right)^{1/2}.
\end{align}

Now we discuss more specific models, which may be subject to stronger constraints.
We consider the following dimension-6 CP-odd operator~\cite{Dine:1990fj},
\begin{align}
  \label{eq:dim-6 violation operator}
  \frac{g^2}{32\pi^2M^2} |H_R|^2 W_R\tilde{W}_R.
\end{align}
The parity symmetry does not forbid this CP-odd operator and only requires another dimension-6 operator composed of $H_L$ and $W_L$ with the same strength.
The operator may be UV-completed by, e.g., the following parity-symmetric interactions and masses,
\begin{align}
  y H_R L' \eta +  \lambda H_R^\dag \bar{L}' \eta + m_L L' \bar{L}' + \frac{1}{2}m_\eta \eta^2 \nonumber \\
  + y^* H_L \bar{L} \eta + \lambda^* H_L^\dag L \eta + m_L^* L \bar{L}  + {\rm h.c.},
\end{align}
with the $SU(2)_L\times SU(2)_R\times U(1)_{X}$ charges of the fermions given by $L'({\bf 1},{\bf 2},-1/2)$, $\bar{L}'({\bf 1},{\bf 2},-1/2)$, $\eta({\bf 1},{\bf 1},0)$, $L({\bf 2},{\bf 1},1/2)$, and $\bar{L}({\bf 2},{\bf 1},1/2)$. The parity symmetry requires that $m_\eta$ is real, but the physical CP phase ${\rm arg}(y \lambda / (m_L m_\eta))$ is in general non-zero.

We make an approximation that the sphaleron rate is given by Eq.~\eqref{eq:sphaleron rate} when $m_{W_R}(H_R) < \sigma \alpha_R T$ and zero for $m_{W_R}(H_R) > \sigma \alpha_R T$. Here $\sigma$ is a constant that can be estimated from the simulation in Ref~\cite{DOnofrio:2014rug}, and we obtain $\sigma \simeq 3$.
Putting all information together, the final baryon-to-entropy ratio is given by
\begin{align}
  \label{eq:ratio}
  Y_B = \frac{n_B}{s} = \left.\frac{28}{79} \frac{1}{s} \frac{4 \Gamma_{\rm sph} \sigma^2 \alpha_R^2 T}{M^2 g^2} \right|_{T=T_n}\simeq 8.9 \times 10^{-11} \left(\frac{1.7 T_n}{M}\right)^2.
\end{align}
The observed baryon asymmetry $Y_B \simeq 8.7 \times 10^{-11}$ can be achieved for
$M$ slightly above
$T_n$.
This justifies the use of the dimension-6 operator; the masses or heavy particles in UV-completion of the operator ($\bar{L}$, $ \bar{L}'$, and $\eta$ in the above example) can be above $T_n$ so that they may be indeed integrated out.
The setup, however, cannot accommodate the dilution necessary for the model with right-handed neutrino dark matter while satisfying $M > T_n$.

In the model with extra scalars,
the following dimension-5 operator can achieve local baryogenesis~\cite{Harigaya:2022ptp},%
\footnote{We assume that the singlets do not couple to the $SU(3)_C$ field strength, since otherwise $\vev{S_L} \neq \vev{S_R}$ generate a non-zero strong CP phase.}
\begin{align}
  \label{eq:dim-5 violation operator}
  \mathcal{L} = \frac{\alpha_R}{8\pi} \frac{S_R W_R \tilde{W_R}}{M}.
\end{align}
This operator may be UV-completed by identifying $S$ with a pseudo Nambu-Goldstone boson with a decay constant $\sim M$ and weak anomaly.
Since the shift of the field value of $S_R$ is around $\Delta S_R \simeq A \Delta h_R^2/(2\mu_S^2)$, the baryon asymmetry is given by Eq.~\eqref{eq:ratio} multiplied by $MA/(2\mu_S^2)$.
Around the allowed parameter space in Fig.~\ref{fig:scalar}, we always have $A/\mu_S \simeq 0.3$. Thus we obtain
\begin{align}
  \label{eq:ratio-scalar}
  Y_B \simeq 8.7\times 10^{-11} \left(\frac{v_R}{20~\tev}\right) \left( \frac{ T_n}{0.2v_R}\right)^2  \left(\frac{40 v_R}{M}\right) \left(\frac{10~\gev}{\mu_S}\right).
\end{align}
Here we normalized $T_n$ by the value we find by computing the bounce action, $T_n \simeq 0.2v_R$.
The smallness of $\mu_S$ causes the large shift of the field value of $S_R$ during the phase transition and strongly enhances the generated baryon asymmetry. As a result, $M$ may be much above $T_n$.

In deriving the parameter space achieving SFOPT by computing the bounce action, we assume $S^2$ and $S|H|^2$ terms and neglect higher-order terms. This requires that the shift of $S$ from the $SU(2)_R$ symmetric point to the escape point in the bounce solution be smaller than $M$, since otherwise higher order terms in general become comparable to the terms we consider. We find that the shift is about $T_n/4$, which is indeed much smaller than $M$.

\section{Signals}
\label{sec:signal}
In this section, we discuss the experimental signals of the model.

\subsection{New gauge bosons}

The left-right symmetric model predicts a new $W_R$ gauge boson and a $Z^{\prime}$ gauge boson. Their masses are given by
\begin{align}
  \label{eq:new W mass}
  m_{W_R}        & = m_W \frac{v_R}{v_L} = 6.5~\tev \frac{v_R}{15~\tev},\nonumber \\
  m_{Z^{\prime}} & = m_Z \frac{v_R}{v_L} = 7.4~\tev \frac{v_R}{15~\tev}.
\end{align}

The new gauge bosons couple with SM leptons and quarks, and thus can be searched for by collider experiments for various final states, including lepton+MET, jets, di-lepton, etc.
The current limit from the $W_R$ search is more stringent than that of $Z^{\prime}$. The $W_R$ boson with a mass below  $6.0~\tev$ is excluded by the search for di-electron final states~\cite{ATLAS:2019lsy}, corresponding to $v_R = 14.1~\tev$. The high-luminosity LHC (HL-LHC) can detect the $W_R$ boson with a mass up to $7.9~\tev$~\cite{ATL-PHYS-PUB-2018-044}, corresponding to $v_R = 18.6~\tev$.

\subsection{New charged particle}

As is discussed in Sec.~\ref{sec:B-L}, a non-zero $B-L$ asymmetry can be produced if $v_R > 10^6~\gev$ or the charged lepton mass has the  structure in Eq.~\eqref{eq:no Dirac mass}. While the former  scenario has no collider signatures, the latter predicts a parity partner of the tau lepton with a mass
\begin{align}
  \label{eq:new lepton mass}
  m_{\tau'} & = m_\tau \frac{v_R}{v_L} \simeq 150~{\rm GeV} \frac{v_R}{15~\tev}
\end{align}
and a hypercharge of unity.
The mass is correlated with the masses of the new gauge bosons $W_R$ and $Z'$.
It is possible that the other two generations of leptons have the structure of the vanishing Dirac mass term $M_{ij}$, for which case even lighter charged leptons are predicted. To be conservative, we consider the case where only $\tau'$ is light
and discuss how to probe it in collider experiments.

$\tau^{\prime}$ is pair-produced by the hypercharge gauge interaction at colliders, whose cross section is computed in, e.g.,~\cite{Kumar:2015tna,Bhattiprolu:2019vdu}.
The decay channel of $\tau^{\prime}$ depends on how the right-handed neutrino masses are obtained.
When they are obtained via a dimension-5 Majorana or Dirac mass operator, $N$ is light so $\tau^{\prime}$ promptly decays into $N$ via an off-shell $W_R$ boson.
The collider signals are then $e^+e^-$, $\mu^+\mu^-$, $e^+ \mu^-$, $e^- \mu^+$, $e + 2$ jets, $\mu + 2$ jets, or 4 jets, all accompanied by missing energy from  right-handed neutrinos.
Note that $\tau$ final states are absent since the SM right-handed tau is $SU(2)_R$ singlet.
The signal resembles that of pair-production of charginos decaying into a neutralino and $W$. Comparing the upper bound on the cross section provided in the HEPData version of~\cite{ATLAS:2019lff} with the production cross section of hyper-charged fermions, we find that there is no LHC constraint. Only the LEP bound of $m_{\tau'} > 100$ GeV~\cite{L3:2001xsz} is applicable.

When the right-handed neutrino mass is generated by Eq.~\eqref{eq:N with chi},
the decay channel depends on whether the right-handed neutrino is heavier than $\tau^{\prime}$.
If $N_3$ is lighter than $\tau^\prime$, $\tau^\prime$ will decay to $N_3$ and emits an off-shell $W_R$ gauge boson.
For $m_{N_3}$ below the EW scale, $N_3$ decays into $\tau$ and an off-shell $W_L$ via the $\nu-N$ mixing $\sim v_L/v_R$ with a typical decay length in the lab frame
\begin{align}
  \label{eq: N lifetime}
  \left( \frac{3m_{N_3}^5}{512 \pi^3 v_L^2 v_R^2} \frac{m_{N_3}}{m_{\tau^\prime}/3}\right)^{-1} \simeq 1~{\rm mm}  \left(\frac{10~\gev}{m_{N_3}}\right)^6 \left( \frac{m_{\tau'}}{200\gev} \right)^3.
\end{align}
For a range of $m_{N_3}$, the decay length is $O(1-10^3)$ mm and $N_3$ can be observed as a displaced vertex. However, since $N_3$ is highly boosted, the decay products of it are collimated and the efficiency of the reconstruction of the displaced vertex may be degraded. The estimation of the bound on $m_{\tau'}$ for this case is beyond the scope of this paper. For sufficiently small $m_{N_3}$, the typical decay length exceeds $O(1)$ m and $N$ is observed as missing energy. The signal is the same as that in the previous paragraph. For $m_{N_3}$ above the EW scale, $N_3$ promptly decays into $\tau + W$ or $\nu + Z/h$. See Sec.~\ref{sec:HNL} for the comment on the search for singlet leptons.

If $N_3$ is heavier than $\tau'$, $\tau^\prime$ decays into an off-shell $N_3$ via $W_R$ exchange, or into an SM fermion and a boson via the Dirac mass term $M_{3} E_3\bar{E}_3$.
We first discuss the former decay mode.
The off-shell $W_R$ at least decays into $ud$ and $sc$. The decay into $t b$ is possible, but this channel is suppressed unless the Dirac mass term for the third generation up quark, $M_{33}^u$, is comparable to $v_R$ so that the SM right-handed top contains a significant fraction of $SU(2)_R$ doublet.
If $N_{1,2}$ are  lighter than $\tau^{\prime}$, decay into $e N_1$ and $\mu N_2$ is possible, and $N_{1,2}$ decays into SM particles as in the previous paragraph.
The off-shell $N_3$ decays into either $\nu$ via $v_L$ insertion, or  into $\nu + Z/h$ or $\tau + W_L$, with $Z/h/W_L$ dominantly decaying into jets or leptons.
For $m_{\tau'} < $ TeV, the former dominates and the decay length of $\tau'$ is given by
\begin{align}
  \left(\frac{1}{1536\pi^3} \frac{m_{\tau'}^5 v_L^2}{v_R^6} N_f\right)^{-1} \simeq 8~ {\rm mm}\times  \frac{m_{\tau'}}{200~{\rm GeV}} \frac{6}{N_f},
\end{align}
where $N_f$ is the number of final states. For example, if only $ud$ and $sc$ are available, $N_f=6$.
The decay length of $\tau'$ is above mm so $\tau'$ can be observed as di-jets from a displaced vertex.
The SM neutrino from the off-shell $N$ leads to missing energy.
The most recent LHC search for displaced vertices provides upper limit for the production cross section~\cite{CMS:2020iwv} as a function of a decay length above $1$ mm. Comparing it with the production cross section of hyper-charged fermions~\cite{Kumar:2015tna,Bhattiprolu:2019vdu}, we obtain a bound $m_{\tau^{\prime}} > 1~\tev$.

On the other hand, $M_{3}$ induces the decay of $\tau^\prime$ into $\tau + Z/h$ or $\nu + W$, and the  decay length is
\begin{align}
  \left(\frac{1}{4\pi} \frac{y_\tau^2 M_{3}^2}{m_{\tau^\prime}}\right)^{-1} \simeq 0.1~\mathrm{mm} \left(\frac{3~\mev}{M_3}\right)^{2} \frac{m_{\tau^\prime}}{200~\gev}.
\end{align}
For $M_{3}$
that saturates the upper bound in Eq.~\eqref{eq:M3 upper bound}, the decay rate is larger than that given by the $W_R$ exchange discussed above and the decay length is below 1 mm; the decay is prompt and only the LEP bound $m_{\tau'}>100$ GeV is applicable. For $M_{3}$ below the MeV scale, the decay length can be long enough that $\tau'$ can be observed as a displaced vertex, for which $m_{\tau'} > 1$ TeV is required.

\subsection{New heavy neutral lepton}
\label{sec:HNL}

In the neutrino mass model in Sec.~\ref{sec:radiative inverse seesaw}, right-handed neutrinos mix with standard model neutrinos with the square of the angle
\begin{align}
  \theta_{\nu N}^2 \simeq \left(\frac{v_L}{v_R}\right)^2 = 3.3\times 10^{-5} \left(\frac{30~{\rm TeV}}{v_R}\right)^2.
\end{align}
If the masses of them are below 100 GeV, right-handed neutrinos are subject to constraints from direct searches. See~\cite{Abdullahi:2022jlv} for the overview of constraints and prospects. This is complementary to the search for a new charged lepton, whose prospects become worse if the right-handed tau neutrino $N_3$ is lighter than $m_{\tau'}$.
Note that $N_3$ is also produced from the decay of hyper-charged $\tau'$; this may help the search for right-handed neutrinos.

\subsection{New scalar particle}
\label{sec:scalar future}
In Sec.~\ref{sec:scalar extension}, we introduced a model with two singlet scalar particles, $S_L$ and $S_R$, and discussed the current experimental limit. While $S_R$ is difficult to probe, $S_L$ can be probed via exotic decay of the SM Higgs at future colliders.
In Fig.~\ref{fig:scalar}, we show the future prospects of HL-LHC and Higgs factory derived in Ref.~\cite{Carena:2022yvx}. The majority of the parameter region can be probed by Higgs factories.

\subsection{Electric dipole moment}

The CP violation to generate baryon asymmetry in general induces electric dipole moments (EDMs) of SM fermions. In the local baryogenesis model in Eq.~\eqref{eq:dim-6 violation operator}, this is dominated by the correction from the parity partner of the operator $\propto |H_L^2|W_L\tilde{W}_L$. The correction is logarithmically enhanced by the renormalization group effect and is given by~\cite{Lue:1996pr}
\begin{align}
  \frac{d_e}{e} \simeq & 1 \times 10^{-30}{\rm cm} \left(\frac{20~\tev}{M}\right)^2 \frac{{\rm ln}(M^2/m_h^2)}{8}\nonumber                                                                        \\
  \simeq               & 1 \times 10^{-30}{\rm cm} \left(\frac{Y_B}{8.9\times 10^{-11}}\right) \left(\frac{20~\tev}{v_R} \right)^2 \left( \frac{v_R}{1.7 T_n}\right)^2 \frac{\ln (M^2/m_h^2)}{8}.
\end{align}
The current bound $d_e/e < 1.1\times 10^{-29}$ cm~\cite{ACME:2018yjb} is satisfied for $v_R$ above the collider bound from $W_R$ search.
Near-future detection of the electron EDM means $v_R = O(10)$ TeV, and $\tau'$ and/or $SU(2)_R$ gauge bosons are predicted to be within the reach of near future collider experiments.

The EDM in the extra scalar model induced by the operator in Eq.~\eqref{eq:dim-5 violation operator} is~\cite{Harigaya:2022ptp}
\begin{align}
  \label{eq:EDM dim-5 result}
  \frac{d_e}{e} \simeq & 1 \times 10^{-31} \mathrm{cm} \left( \frac{20~\tev}{v_R} \right) \left( \frac{40 v_R}{M} \right) \left( \frac{\mu_S}{10~\gev} \right) \nonumber                                               \\
  \simeq               & 1 \times 10^{-31} \mathrm{cm} \left( \frac{Y_B}{8.7\times 10^{-11}} \right) \left( \frac{20~\tev}{v_R} \right)^2 \left( \frac{0.2 v_R}{T_n} \right)^2 \left( \frac{\mu_S}{10~\gev} \right)^2,
\end{align}
where we use Eq.~\eqref{eq:ratio-scalar} in the second equality.
The viable parameter region with $\mu_S\sim 10$ GeV can be probed by future electron EDM measurements~\cite{Vutha:2009ux} if $v_R= O(10)$ TeV.

In the model with right-handed neutrino dark matter, with the dilution in Eq.~\eqref{eq:dilution}, the required $M$ becomes smaller and the prediction on the EDM becomes larger,
\begin{align}
  \frac{d_e}{e}
  \simeq  1 \times 10^{-30} ~\mathrm{cm} \left( \frac{\mu_S}{10~\gev} \right)^2 \left(\frac{\sum_i m_{\nu_i}}{60~\mathrm{meV}} \right) \left(\frac{80}{g_s(T_D)}\right) \left( \frac{Y_B}{8.7\times 10^{-11}} \right) \left( \frac{0.2 v_R}{T_n} \right)^2.
\end{align}
Interestingly, the prediction is independent of $v_R$ and is within the reach of near-future measurements.

\subsection{Gravitational waves}

In order for the gravitational waves (GWs) produced by the $SU(2)_R$ phase transition to be observable, a very strong phase transition is required.
A marginally SFOPT with $\vev{H_R}\sim T$ at the nucleation temperature is not enough, and high-temperature expansion is not justified for the analysis of very SFOPT. We thus use the full form of the 4d effective potential in Appendix~\ref{sec:potential} to numerically compute the gravitational wave signal.
The numerical computation shows that the GW signals are suppressed in the models using the running quartic coupling or extra light singlet scalars discussed in the previous sections. Here we make the following simple arguments instead of showing a detailed computation.
\begin{itemize}
  \item For the minimal setup using the running quartic coupling, the SFOPT is achieved only for $v_R \simeq 2 \times 10^8~\gev$. Such a high $v_R$ will make the GW-signal peak at a very high frequency, which is beyond the sensitivity of future observations.
  \item For the running of the quartic coupling assisted by extra leptons, the SFOPT can be achieved either by a relatively higher $v_R$ or a smaller $v_R$ but with a large Yukawa coupling with the extra lepton to speed up the running. For the former case, the typical frequency of GWs is too high. For the latter, the large Yukawa coupling itself also suppresses the strength of the phase transition and makes it only marginally SFOPT or even cannot achieve a SFOPT. The resultant GW signal is weak and cannot be detected.
  \item For the model with a light singlet scalar, the scalar is very light compared with the $h_R$ field. This leads to a huge kinetic energy of $S_R$ during the $SU(2)_R$ phase transition. Such a huge kinetic energy makes the duration of the phase transition short (i.e., a large $\beta/H$ parameter,) and thus suppresses the GW signal~\cite{Harigaya:2022ptp}. Details can be found in Appendix~\ref{sec:analysis-light-scalar}.
\end{itemize}

In other models to realize a SFOPT, the GW signal may be observable. In such a model, the GW signal will be correlated with $v_R$, and hence with the new gauge boson and fermion masses and the EDM.

\section{Summary and discussion}
\label{sec:summary}

The parity solution to the strong CP problem introduces the $SU(2)_R$ gauge symmetry.
In this paper, we proposed a model of baryogenesis from a first-order $SU(2)_R$ phase transition.

The key ingredient of the model is how to obtain a non-zero $B-L$ asymmetry of the SM particles. Although the $SU(2)_R$ sphaleron process does not produce a non-zero charge of the naive $B-L$ symmetry with $\bar{q}(-1/3)$ and $\bar{\ell}(+1)$, the washout by the $SU(2)_L$ sphaleron process is avoided if some of the asymmetry of $SU(2)_R$-charged particles is (temporarily) not transferred into that of $SU(2)_L$-charged particles. This is indeed possible in the model with the minimal Higgs content $H_L$ and $H_R$, where the strong CP problem is solved without introducing extra symmetries.
The scheme predicts a new hyper-charged fermion whose mass is correlated with the $SU(2)_R$ symmetry-breaking scale $v_R$.

First-order $SU(2)_R$ phase transition can be realized by the coupling of Higgses to other fields, as in electroweak baryogenesis. We studied two models that do not introduce extra hierarchy problems beyond that of $H_R$ and $H_L$.
We considered the running of the quartic coupling. If the running of the SM Higgs coupling is the SM one, $v_R>10^8$ GeV is required. With Yukawa couplings to extra fermions, $v_R$ can be much lower. We also considered a coupling of the Higgses to singlet scalar fields. $v_R$ can be as low as the experimental lower bound from $W_R$ search. The scalar mass is predicted to be around $10$ GeV and the scalar can be probed by future colliders via its mixing with the SM Higgs.

CP violation may be introduced in various ways.
We applied the idea of local electroweak baryogenesis to $SU(2)_R$ phase transition and studied the dimension-6 coupling of the $SU(2)_R$ gauge field with $H_R$  and the dimension-5 coupling with a singlet scalar. The higher dimensional operators generate a non-zero electron EDM that is observable in near future experiments if $v_R = O(10)$ TeV. It will be interesting to investigate other realizations of CP violation and the associated predictions on EDMs.

An obvious drawback of the model in comparison with the electroweak baryogenesis is that we generically cannot predict the scale of new physics. Indeed, the model works even if $v_R \gg 100$ TeV, for which observable signals of the model in near future experiments are not guaranteed. However, the model still has several correlated observable signals if $v_R$ is small enough and hence maintains predictability in a loose sense.
Also, in the model with right-handed neutrino dark matter, $v_R$ is bounded from above and the mass of a new hyper-charged fermion is predicted to be below $1$ TeV. It will be worth considering other models with dark matter candidates and making predictions on the energy scale of parity-symmetric models.

\section*{Acknowledgement}
I.R.W is supported by DOE grant DOE-SC0010008. We thank Philipp Schicho for useful discussions about the usage of \verb|DRalgo| package.

\appendix

\section{Uncertainties in $V_{\rm eff}$ computation: the Standard Model with light Higgs and top}
\label{sec:potential}

The computation of the thermal effective potential
involve uncertainties, which may introduce corresponding uncertainties
in the prediction on the boundary of SFOPT, i.e., $\phi_c/T_c = 1$~\cite{Croon:2020cgk,Schicho:2021gca,Niemi:2021qvp,Schicho:2022wty}. In this appendix, we summarize different computation methods and compare their results. Since all the SFOPT approaches mentioned in this paper are intrinsically the same as the Standard Model with a light Higgs, to make the comparison with the literature easier,  we parameterize the gauge and top Yukawa couplings by the masses of particles with the symmetry breaking scale at the vacuum being the electroweak scale, i.e., $173$ GeV. As is found in~\cite{Dine:1992wr}, the phase transition strength is very sensitive to the top Yukawa coupling.
To analyse the
$SU(2)_R$ phase transition that occurs at a much higher temperature than the electroweak scale,
where the top Yukawa is smaller, we choose two much smaller top Yukawa couplings that correspond to $m_t = 120$ and $ 90$ GeV, and slightly smaller electroweak gauge coupling constant that corresponds to $m_W = 73$ GeV, $m_Z = 83$ GeV.

At 1-loop level, a traditional computation gives the effective potential (for early papers and good reviews, see~\cite{Coleman:1973jx,Jackiw:1974cv,Kang:1974yj,Dolan:1974gu,Fukuda:1975di,Aitchison:1983ns,Dine:1992wr,Arnold:1992rz,Loinaz:1997td,Quiros:1999jp,Morrissey:2012db,Croon:2020cgk})
\begin{align}
  \label{eq:Veff}
  V(h,T) = V_{T=0} + V_{\rm FT},
\end{align}
where
\begin{align}
  \label{eq:thermal potential}
  V_{T=0}    & = - \frac{1}{2}\mu_{h}^2 h^2 + \frac{1}{4}\lambda h^4 + 4 B v^2 h^2 - \frac{3}{2} B h^4 + B h^4 \ln \left( \frac{h^2}{2 v^2} \right),\nonumber \\
  B          & =  \frac{1}{256\pi^2 v^4} \left(2m_W^4 + m_Z^4- 4m_t^4\right),\nonumber                                                                          \\
  V_{\rm FT} & = \frac{T^4}{2\pi} \left( 4 J_B(\frac{m_{W}}{T}) + 2 J_B(\frac{m_Z}{T}) - 12 J_F(\frac{m_t}{T}) \right)\nonumber \\
  J_{B,F}(r^2) &= \int_0^\infty dx x^2 \log \left( 1 \mp \exp(-\sqrt{x^2 + r^2}) \right).
\end{align}
Here we use the on-shell renormalization scheme.
To deal with the infrared divergence of the zero-mode of bosonic degree of freedom, proper resummation is needed. The resummation, however, is still controversial, and different resummation method causes uncertainties. A commonly used method is to add a ring-diagram contribution $V_{\rm ring}$ to Eq.~\eqref{eq:thermal potential} (the Arnold-Espinosa resummation), i.e.,
\begin{align}
  \label{eq:ring}
  V(h,T)       & = V_{T=0} + V_{\rm FT} + V_{\rm ring},\nonumber                                                                         \\
  V_{\rm ring} & = - \frac{T}{12\pi} \left( 2 (m_W(h,T)^{3/2} - m_W(h)^{3/2}) + (m_Z(h,T)^{3/2} - m_Z(h)^{3/2} )+ m_B(h,T)^{3/2}\right),
\end{align}
where the temperature-dependent masses are the thermal masses and can be found in~\cite{Carrington:1991hz}.

Around the boundary of SFOPT, we have $T > m(h)$ for all the degrees of freedom and we may use the high-temperature expansion
\begin{align}
  \label{eq:highT V}
  V & = D (T^2 - T_0^2)h^2 - E T h^3 + \frac{1}{4}\lambda(T)h^4,
\end{align}
where
\begin{align}
  \label{eq:highT coefficients}
  D          & = \frac{1}{16 v^2}\left (2m_W^2 + m_Z^2 + 2m_t^2\right ),~
  E           = \frac{2}{3}\frac{1}{8\sqrt{2}\pi v^3} \left(2m_W^3 + m_Z^3 \right),~
  T_0^2       = \frac{1}{2D} (\mu_{h}^2 - 4B v^2), \nonumber                                                                                                           \\
  \lambda(T) & = \lambda - \frac{3}{64 \pi^2 v^4}\left(2m_W^4 \ln \frac{m_W^2}{a_B T^2} + m_Z^4 \ln \frac{m_Z^2}{a_B T^2} - 4 m_t^4 \ln \frac{m_t^2}{a_F T^2} \right).
\end{align}
Here $\log a_B = 2 \log 4\pi - 2\gamma_E$ and $\log a_F = 2\log \pi - 2\gamma_E$, with $\gamma_E$ the Euler-gamma. Here we add a factor of $2/3$ in the $E$ term to screen out the longitudinal instead of adding the ring-diagram term. We can further simplify the expression by neglecting the log terms, which leads to the most simplified expression
\begin{align}
  \label{eq:highT simple}
  V     & = D (T^2 - T_0^2)h^2 - E T h^3 + \frac{1}{4}\lambda h^4,\nonumber                \\
  D     & = \frac{1}{16 v^2}\left (2m_W^2 + m_Z^2 + 2m_t^2\right ),~
  E      = \frac{2}{3}\frac{1}{8\sqrt{2}\pi v^3} \left(2m_W^3 + m_Z^3 \right),~
  T_0^2  = \frac{1}{2D} \mu_{h}^2 .
\end{align}
The phase transition strength $\phi_c/T_c $ can be simply expressed as $2E/\lambda$ under this expression, and
we can see that a smaller quartic coupling $\lambda$ leads to large phase transition strength.
Even for the cases where the high-temperature expansion is not a good approximation, this still provides a useful intuition.

The main problem of Eq.~\eqref{eq:highT simple}, however,  is its complete ignorance of the zero-temperature correction. From Eq.~\eqref{eq:highT coefficients} where such an effect is included, we can see that the $\lambda$ receives a positive contribution from the top Yukawa coupling, i.e. $m_t$, which suppresses the phase transition strength.
Such a result is confirmed by numerical computation from the full form.
The suppression of the phase transition strength by the top Yukawa is relieved at a high $v_R$, since the top Yukawa coupling is much smaller than that at the EW scale due to running.

\begin{figure}[h]
  \label{fig:thermal potential}
  \centering
  \includegraphics[width=0.47\linewidth]{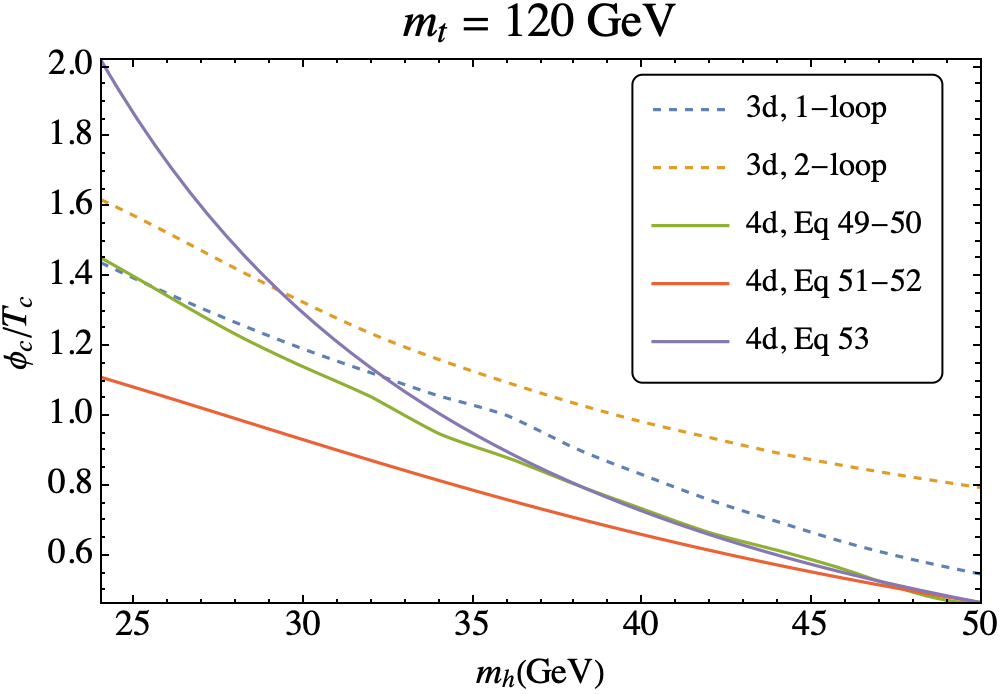}
  \includegraphics[width=0.47\linewidth]{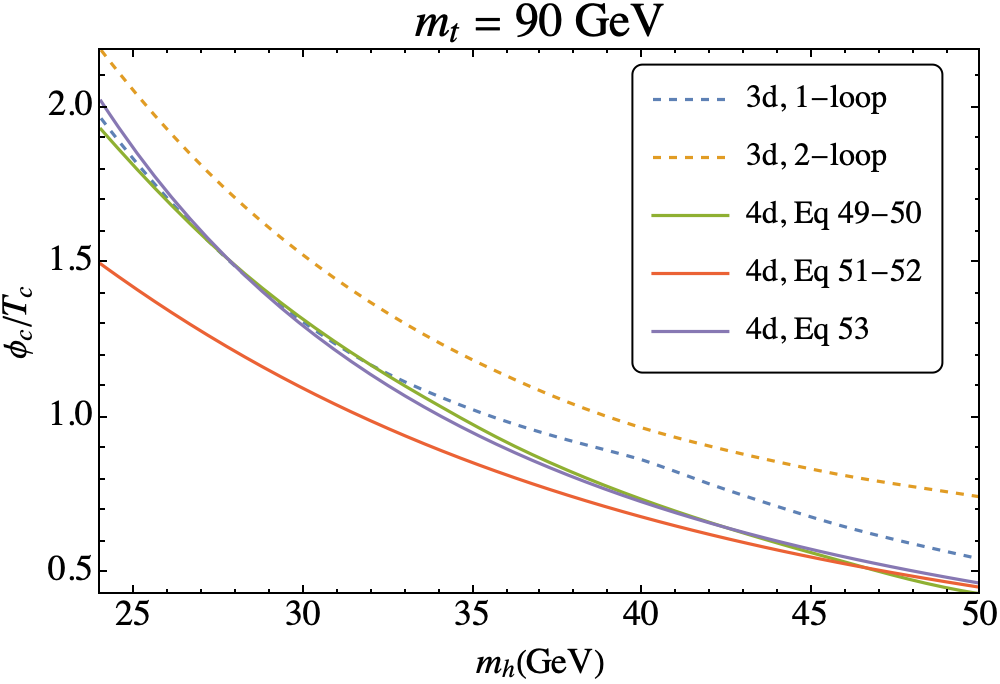}
  \caption{The phase transition strength $\phi_c/T_c$ under different computation methods. Dashed lines are computed in the 3d effective theory, while the solid lines are performed in the 4d theory. The 4d theories are computed under the full form (Eq~\eqref{eq:thermal potential}-\eqref{eq:ring}), or high-temperature expansion with (Eq~\eqref{eq:highT V}-\eqref{eq:highT coefficients}) or without (Eq~\eqref{eq:highT simple}) the leading log terms.}
\end{figure}

On the other hand, the state-of-the-art method to perform the thermal resummation is to integrate out the zero mode of the bosonic degrees of freedom and work in an effective dimension-3 theory, so-called the dimensional reduction~\cite{Ginsparg:1980ef,Appelquist:1981vg,Nadkarni:1982kb,Landsman:1989be,Kajantie:1995dw}. This computation method is theoretically more reliable while also being more complicated. With the help of the recently published \verb|Mathematica| package \verb|DRalgo|~\cite{Ekstedt:2022bff}, we compute the thermal potential up to 2-loop level. In Fig~\ref{fig:thermal potential}, we show the comparison of different computation methods. 
One can see that the agreement between different approximation methods gets better for smaller $m_t$.
Since scanning over the parameter space using the dimensional reduction with 2-loop correction is complicated, we use the approximation in  Eq.~\eqref{eq:highT simple} to derive the boundary of the SFOPT in the main text because of its simplicity while being in reasonable agreement with the state-of-the-art 2-loop computation result.

\section{Analysis of the light scalar model}
\label{sec:analysis-light-scalar}

In Sec.\ref{sec:scalar extension}, we proposed a model with a naturally light scalar to enhance the phase transition. In this appendix, we discuss some important features of this model, including the fine-tuning of the scalar mass and the origin of large $\beta/H$, which is relevant to the discussion of the gravitational-wave signals.
As in the previous section,
we analyze the SM with 
a lighter top quark mass $m_t = 120~\gev$.
The Higgs boson and gauge boson masses are chosen to be the Standard Model values at EW scale.

The tree-level potential can be written as~\cite{Das:2009ue,Harigaya:2022ptp}
\begin{align}
  V_0 = - \frac{1}{2} \mu_h^2 h^2 + \frac{1}{4}\lambda h^4 + \frac{1}{2}\mu_S^2 S^2 - \frac{1}{2}A S (h^2 - 2 v^2),
\end{align}
where $h$ is the SM Higgs field, $v \simeq 174~\gev$ is the electroweak vev, and $S$ is an extra scalar. 
An approximate shift symmetry $S \rightarrow S + \delta S$, softly broken by the mass term and the trilinear coupling $A$, avoids an extra hierarchy problem; the lightness of $S$ is natural~\cite{Harigaya:2022ptp}. Higher order terms of $S$ is forbidden by the shift symmetry. 

For a given field value of $h$, we may minimize the potential with respect to $S$,
\begin{align}
  \label{eq:Svev-tree}
  \langle S \rangle =& \frac{A}{2 \mu_S^2} h^2 + \text{const}, \nonumber \\
  V_0 =& - \frac{1}{2} \mu_h^2 h^2 + \frac{1}{4}(\lambda - \frac{A^2}{2\mu_S^2}) h^4.
\end{align}
One can see that along this path, the effective Higgs quartic coupling $\lambda$ is corrected to be $\lambda_{\rm eff} = \lambda - A^2/2 \mu_S^2 < \lambda$, which can induce a strong first-order phase transition.

The physical mass eigenstates are
\begin{align}
  \label{eq:physical scalar masses}
    m_+^2 \equiv m_h^2 \simeq & 4 \lambda v^2 (1+ \frac{A^2}{8 \lambda^2 v^2}), \nonumber \\
    m_-^2 \equiv m_S^2 \simeq & \mu_S^2  - \frac{A^2}{2 \lambda},
\end{align}
The full tree-level relationship between parameters can be found in ref~\cite{Harigaya:2022ptp}. The fine-tuning of this model can be defined as
\begin{align}
  \text{fine-tuning} \equiv \frac{\lambda - A^2/2 \mu_S^2}{\lambda} = \frac{m_S^2}{\mu_S^2}.
\end{align}

\begin{figure}[!t]
  \centering
  \includegraphics[width=0.6\linewidth]{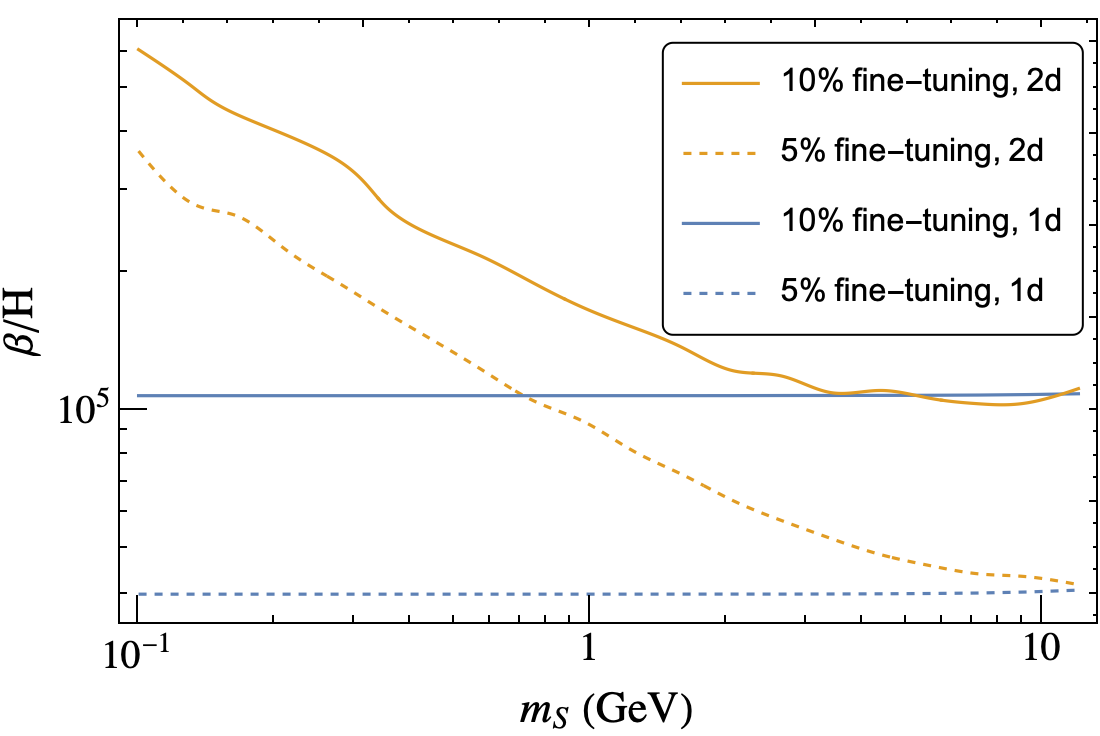}
  \caption{$\beta/H$ as a function of $m_S$ for fixed fine-tuning $5\%$ and $10\%$. 1d: The bounce action is obtained along the path in Eq.~(\ref{eq:Svev-tree}) with the kinetic term of $S$ neglected. 2d: Full two-field dynamics is included.}
  \label{fig:beta-H plot}
\end{figure}

From Eq.~\eqref{eq:Svev-tree}, one can see that the field shift of $S$ is very large for a light $\mu_S$ during the phase transition, if the phase transition path is indeed along that path. Numerical solution to the bounce equation shows that it is indeed the case. The gradient term of the $S$ field, $K_S$, is much larger compared to that of the Higgs field, $K_h$;
\begin{align}
  \label{eq:kinetic energy ratio}
  \frac{K_S}{K_h} \equiv \left (\frac{d S}{d r} \right )^2 /\left( \frac{d h}{d r} \right)^2 = \frac{A^2}{\mu_S^2} \frac{h^2}{\mu_S^2} = 2 \lambda \left( 1 - \text{fine-tuning}\right) \frac{h^2}{\mu_S^2} \simeq 2 \lambda \left( 1 - \text{fine-tuning}\right) \frac{v^2}{\mu_S^2} ,
\end{align}
where $r$ is the space variable along the bubble profile, $S(r)$ and $h(r)$ are the singlet and Higgs field along the bubble profile, respectively. The kinetic energy $E_{k_i}$ is defined as the integral of $r^2 K_i/2$ along the bubble profile. From Eq.~\eqref{eq:kinetic energy ratio}, one can see that a large $v/\mu_S$ ratio leads to large kinetic energy of $S$.
As the temperature cools down, this large, positive contribution to the 3d Euclidean action $S_3$ is canceled by the negative contribution from the potential energy so that $S_3/T \simeq 140$ is finally reached for bubbles to efficiently nucleate. However, the cancellation does not occur for the derivative of $S_3/T$, causing a large $\beta/H \equiv T d(S_3/T)/dT$ at the nucleation temperature.

In Fig.~\ref{fig:scalar}, we show $\beta/H$ as a function of $m_S$ for fixed fine-tuning $5\%$ and $10\%$.
For the orange lines labelled ``2d", we obtain the bounce action with the full two-field dynamics of $h$ and $S$.
One can easily see that $\beta/H$ increases for smaller $\mu_S$. In the parity-symmetric model, the relevant ratio $v_R/\mu_S$ is order-of-magnitude larger than $v/\mu_S$ in the electroweak case in the allowed parameter space in Fig.~\ref{fig:scalar}, so we expect that the $\beta/H$ quantity is even larger than that in Fig.~\ref{fig:beta-H plot}, which suppresses the gravitational wave signal.
For a comparison, we also obtain the bounce action along the path in the first line of Eq.~\eqref{eq:Svev-tree} ignoring the kinetic term of $S$, which corresponds to the one-field dynamics of $h$ with the potential in the second line of Eq.~\eqref{eq:Svev-tree}. The resultant $\beta/H$ is shown by blue lines in Fig.~\ref{fig:beta-H plot}. For large $m_S$, the result is in good agreement with the result of the full two-field computation, since the field excursion of $S$ is small and the kinetic term of $S$ is negligible. For small $m_S$, the kinetic term of $S$ is not negligible and one should solve the full two-field dynamics.

\bibliographystyle{utphys}
\small{\bibliography{Local_EWBG_Parity}}

\end{document}